\documentclass[useAMS,usenatbib]{mnras} 

\usepackage{graphicx}   
\usepackage[english]{babel}

\usepackage{graphicx}   
\usepackage{threeparttable}

\newcommand{\SII}{[S~{\sc ii}]}
\newcommand{\OIII}{[O~{\sc iii}]}

\newcommand{\NII}{[N~{\sc ii}]}
\newcommand{\HII}{H~{\sc ii}}
\newcommand{\HI}{H~{\sc i}}
\newcommand{\Ha}{H$\alpha$}
\newcommand{\Hb}{H$\beta$}
\newcommand{\km}{\,\mbox{km}\,\mbox{s}^{-1}}

\newcommand{\SIIHa}{[S~{\sc ii}]/H$\alpha$}
\newcommand{\NIIHa}{[N~{\sc ii}]/H$\alpha$}
\newcommand{\OIIIHb}{[O~{\sc iii}]5007/H$\beta$}

\title[HII versus HI in `green valley' galaxies ]{\HII\ versus \HI\ in the `green valley' galaxies: direct comparison}

\author[Bryukhareva \& Moiseev]{
  Tatiana S. Bryukhareva$^1$\thanks{E-mail: ts.bryukhareva@physics.msu.ru}, 
    Alexei V.~Moiseev$^{2}$\thanks{E-mail: moisav@gmail.ru}
    \\
 $^{1}$ Lomonosov Moscow State University, Sternberg Astronomical Institute,
        Universitetsky pr. 13, Moscow 119234, Russia
              \\
 $^{2}$ Special Astrophysical Observatory, Russian Academy of Sciences, Nizhny Arkhyz 369167, Russia       
}       

\date{Accepted 2019 Month 00. Received 2019 Month 00; in original form 2017 Month 00}

\begin{document}
\label{firstpage}
\pagerange{\pageref{firstpage}--\pageref{lastpage}} \pubyear{2016}
\maketitle

\begin{abstract}
We study the morphology and kinematics of the ionization state of the gas in four `green valley' early-type galaxies at different stages of their transition from a `blue cloud' of star-forming galaxies to the sequence of passive evolution. The previous \HI\ mapping of the considered sample reveals a spatial offset between the cold gas reservoirs and stellar discs depending on the post-starburst age. Consideration of the ionized-gas properties is essential to understand the role of various feedback processes in star formation quenching. We performed long-slit and 3D optical spectroscopic observations at the 6-m Russian telescope and compared the gas and stellar kinematics. Spatial distribution of the ionized gas is in agreement with \HI\ maps; however, the one-order higher angular resolution in the \HII\ velocity fields allows us to study non-circular gas motions in detail, like the AGN-driven outflow in the nucleus of J1117+51. The most intriguing result is the global \HI+\HII\ gas counter-rotation relative to the stellar disc  in J1237+39. Therefore, in this case all the observed gas in the  `green valley' galaxy was captured from the environment via accretion or minor merging.

\end{abstract}


\section{Introduction}

The optical colours of local galaxies  demonstrate  the well--known bimodality  \citep{Strateva,Baldry04,Baldry06}. Most galaxies can be related either to the `blue cloud' populated with disc-dominated galaxies possessing star-forming cold gas  and young stellar population, or to the `red sequence' -- early-type galaxies (ETG,  ellipticals and lenticuliar) which have exhausted gas reservoirs and contain older stellar population \citep{Blanton09}. An intermediate  `green valley' population consists of galaxies that, as is believed, undergo transitional processes. Traditionally the transition path is considered to lead from the `blue cloud' to the `red sequence'.  As the population of the `green valley' is rather small compared to red and blue groups, star formation quenching  should occur within a relatively short time-scale \citep{2007ApJS..173..342M}.

A place of lenticular galaxies with green optical colours in galaxy evolution is still uncertain. We do not know if these disc galaxies precede gas-rich spirals, have accumulated cold  gas from the environment  at further evolution stages, or, conversely, if they are the former spirals which have lost their gas reservoirs and transited to the red sequence. Both paths have been widely discussed and approved by different studies,  see, for instance, \citet{Silchenko2012,Katkov2015} for the first scenario and \citet{Fasano2000,Wilman2009} for the second. Recently, new observational arguments in favour of the first scenario were presented by \citet*{Silchenko2019}.

Major mergers also are a feasible explanation for the transformation of spirals into lenticulars  \citep{Querejeta2015}. 
Available observations presume contradictory interpretation frequently, that is why it is crucial to use multiwavelength data to understand the balance between   different  gas feedback mechanisms (supernova explosions, active galactic nuclei, galactic winds) and external accretion processes \citep{Fraternali2008}.



\begin{figure*}
\centerline{\includegraphics[height=0.3\linewidth]{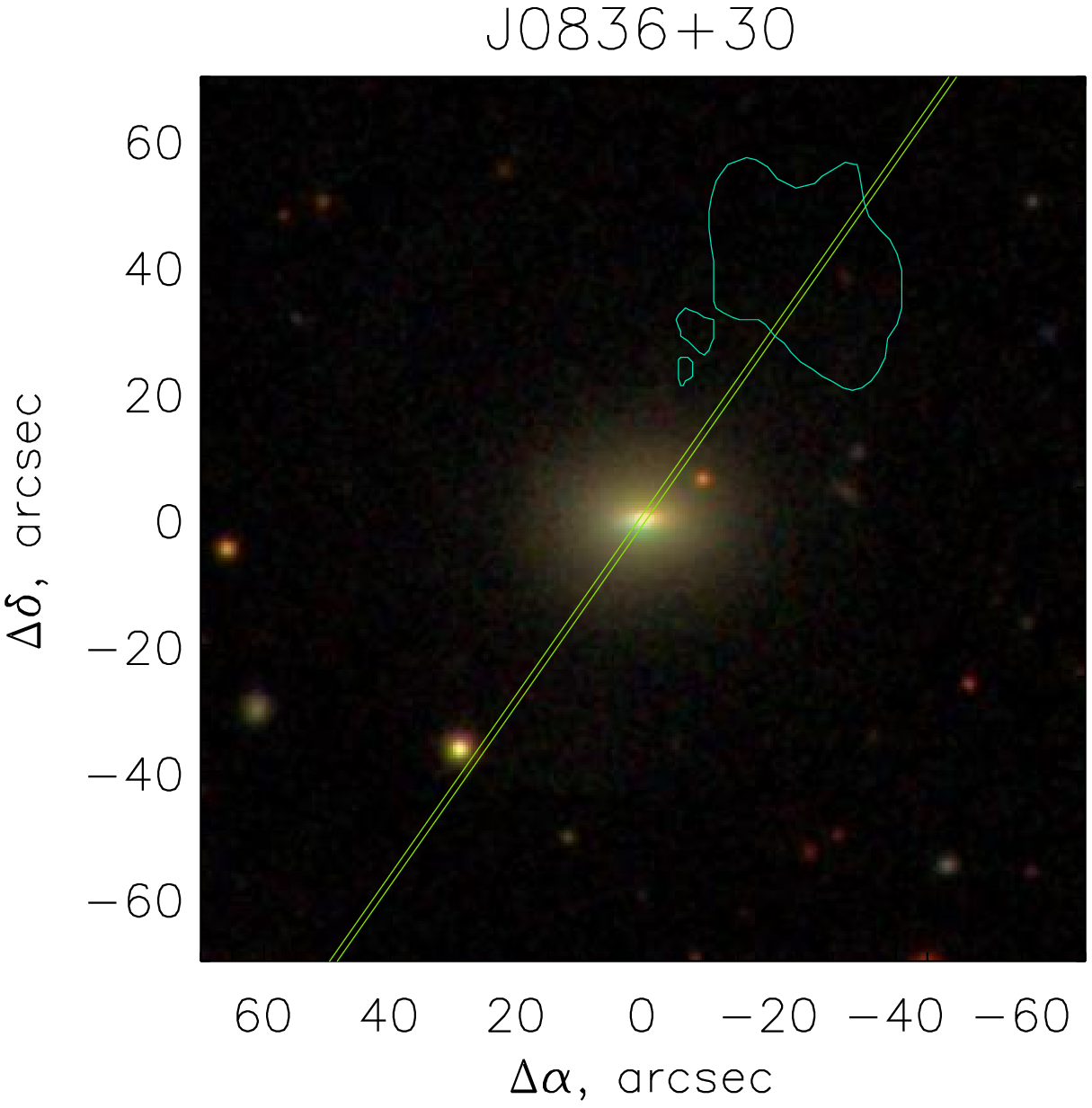}
\includegraphics[height=0.3\linewidth]{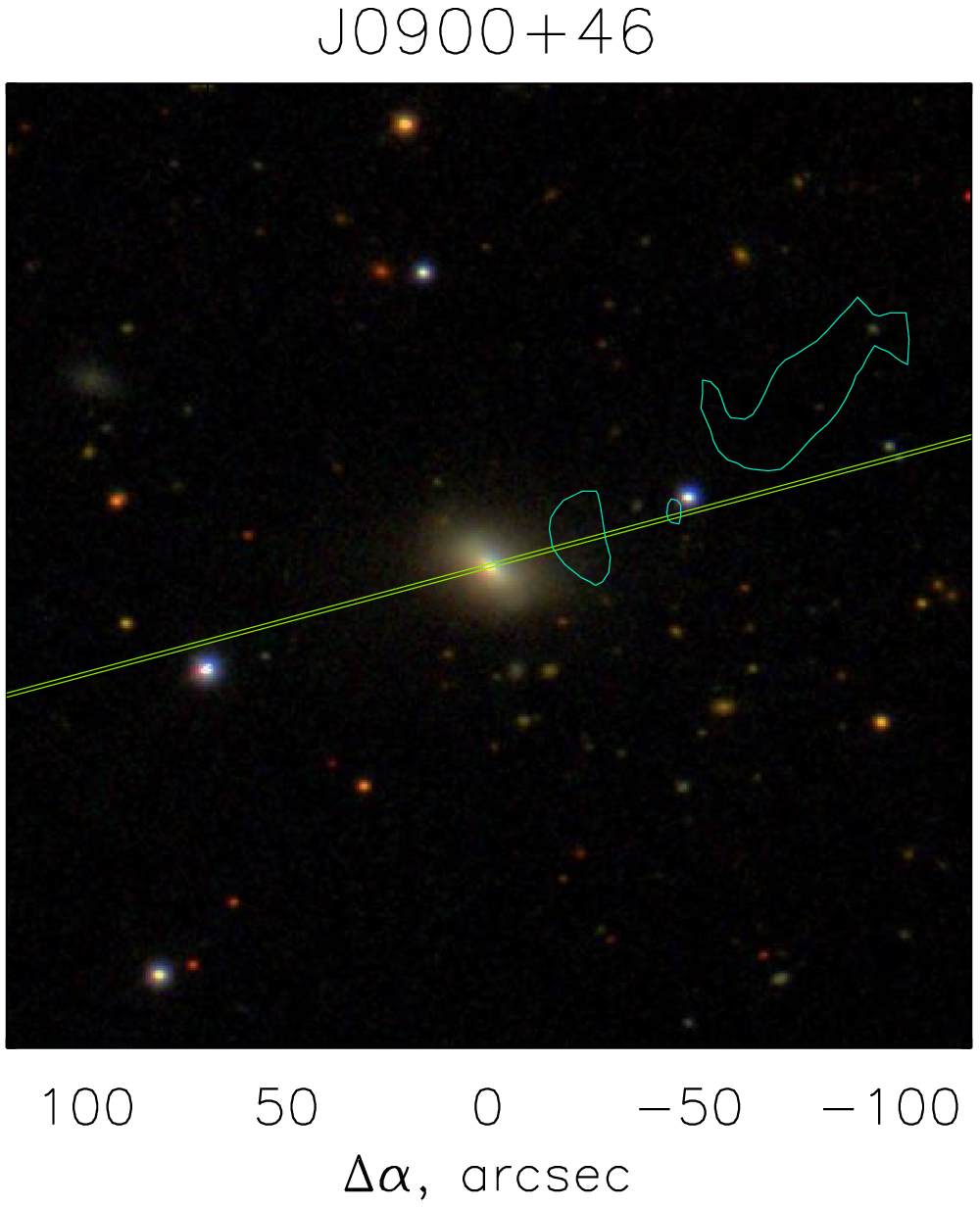}
\includegraphics[height=0.3\linewidth]{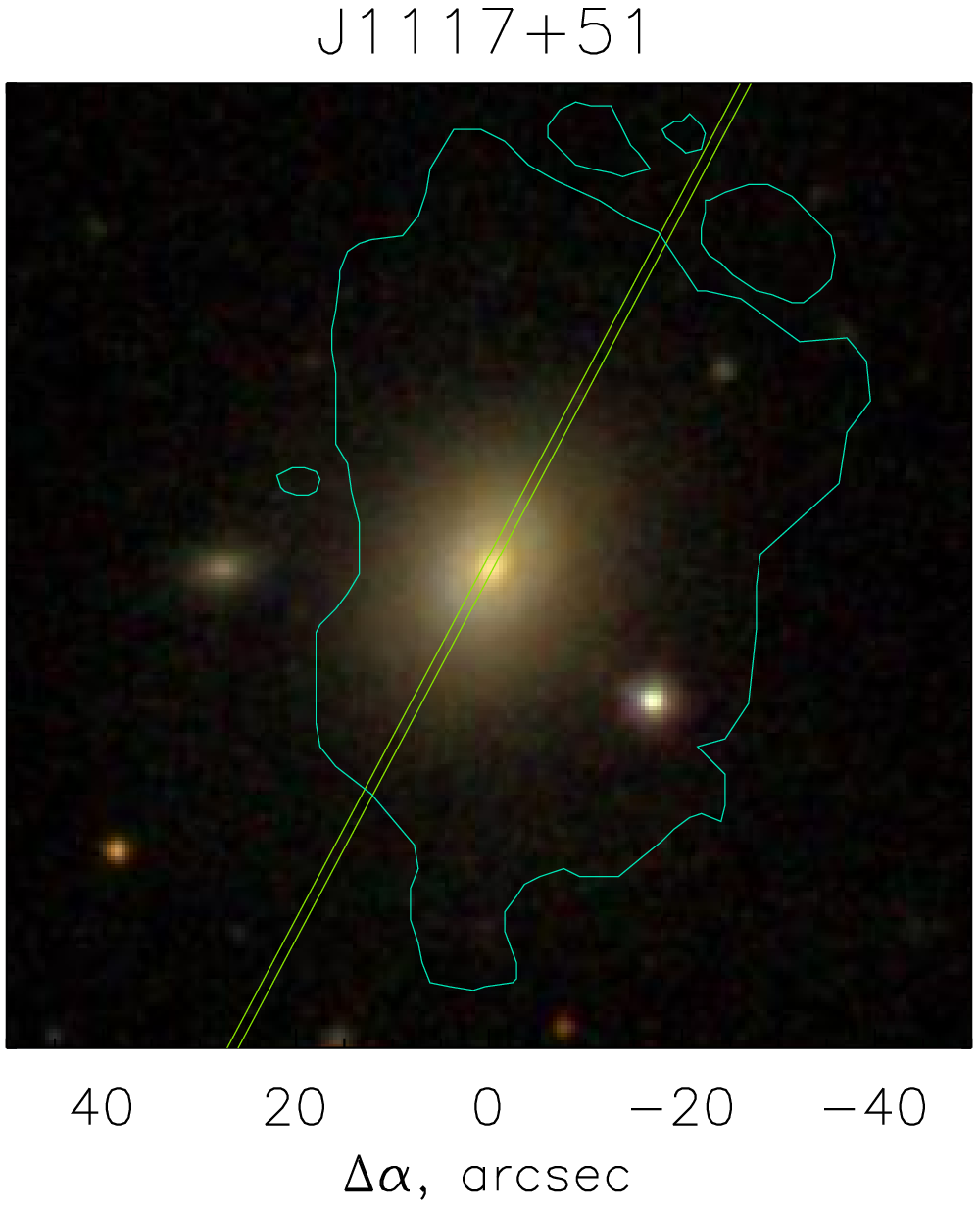}
\includegraphics[height=0.3\linewidth]{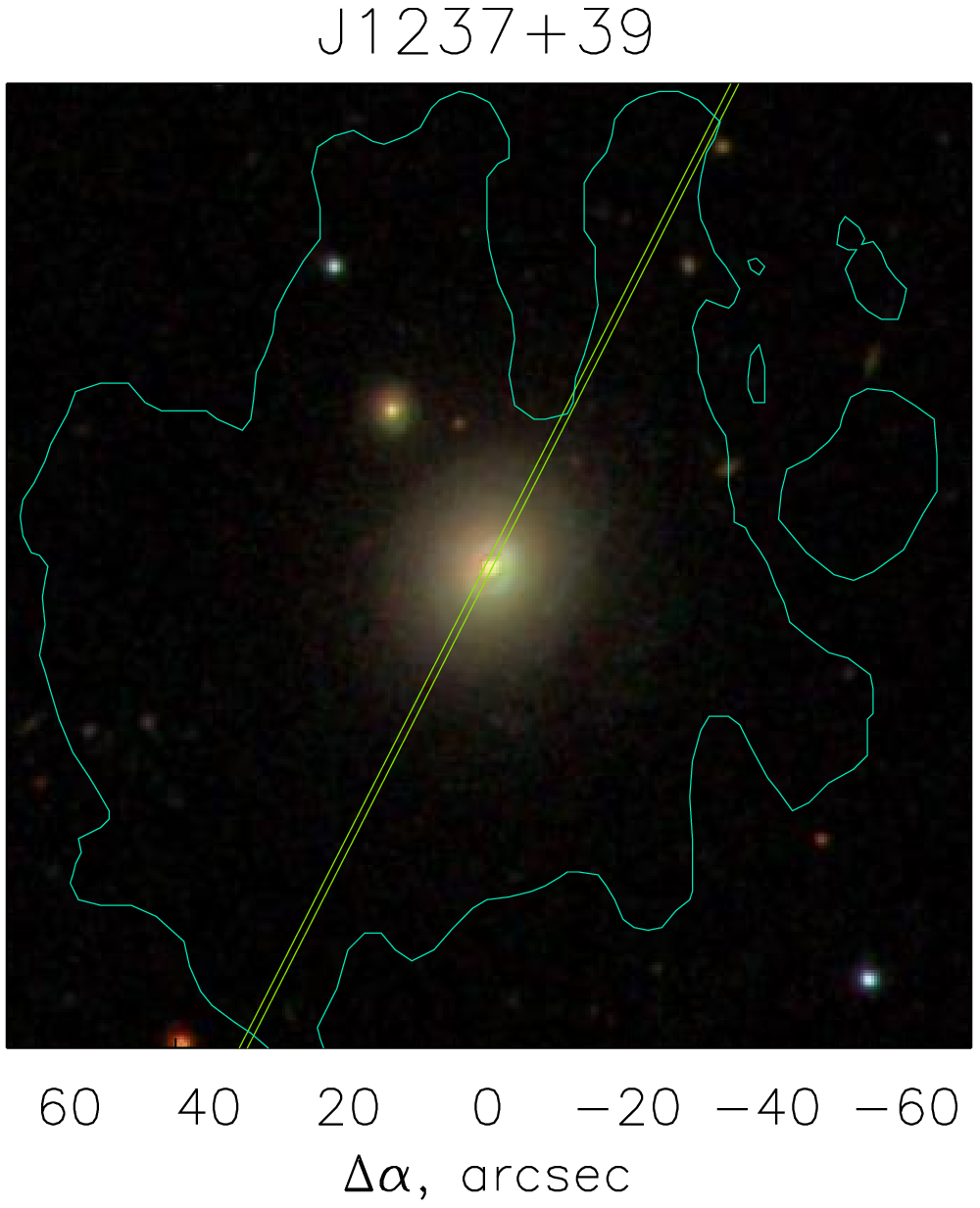}}
\caption{SDSS DR13 images of the observed galaxies. Position of the SCORPIO-2 spectrograph slit is shown in green. The cyan contours mark the external borders  of the \HI\ structures observed with WSRT \citep{Wong2015}. }
\label{fig:sdss}
\end{figure*}

In \citet{Wong2015}, a sample of so-called `blue ETGs', located in the `green valley', was considered. Using the Westerbork Syntheses Radio Telescope (WSRT) they mapped the \HI\ in four local (redshifts $0.02<z<0.03$) galaxies with comparable stellar masses and similar lenticular-like morphology. Their near-ultraviolet colours, however, range from very blue, consistent with the ongoing star formation (J1237+39), to red colours of passive quenched galaxies (J0836+30). 
The coordinates and properties of the considered galaxies are listed in Table~\ref{tab:sample}. Fig.~\ref{fig:sdss} shows the HI distribution with the cyan contours overlaid on optical images. In the galaxies with the bluest NUV colours, undisturbed (J1237+39) and slightly asymmetric (J1117+51) discs are observed. As for the reddest galaxies, J0900+46 and J0836+30, all the cold gas reservoirs are displaced from the stellar discs.

\begin{table*}
	\centering
\caption{Properties of the galaxy sample.}
\label{tab:sample}
    \begin{tabular}{|ccccccl|} 
    \hline
    Galaxy & RA (J2000) & Declination (J2000) & Distance$^a$, Mpc & log(M$_*$)$^a$ & lg M$_{HI}^a$ & \HI\ summary$^a$  \\ 
    \hline
    J0836+30 & 08:36:01.5 & +30:15:59.1 & 105 & 10.7 & 8.0 & \HI\ ejected \\
    J0900+46 & 09:00:36.1 & +46:41:11.4 & 113 & 10.5 & 9.1 & \HI\ ejected \\
    J1117+51 & 11:17:33.3 & +51:16:17.7 & 115 & 10.6 & 8.5 & \HI\ displaced \\
    J1237+39 & 12:37:15.7 & +39:28:59.3 & 84 & 10.3 & 9.6 & \HI\ rotating disc \\
    \hline
    \end{tabular}
    \begin{tablenotes}
      \item $^a$ From \citet{Wong2015}.
    \end{tablenotes}
\end{table*}

\begin{table*}
	\centering
\caption{Log of SCORPIO-2 observations.}
\label{tab:obs}
\begin{tabular}{|ccccrcc|} 
\hline
  Galaxy & Date of obs. & SCORPIO-2 mode & Sp.range, \AA&  Exp. time, s &  Seeing, $''$ & Slit PA, deg.\\ 
\hline
J0836+30 & 07.12.2015 & Long-slit & 3600--7070 & $4\times900$ &   1.9 & 145 \\
J0900+46 & 05.12.2015 & Long-slit & 3600--7070 & $6\times600$ &   2.2 & 105 \\
J1117+51 & 08.12.2015 & Long-slit & 3600--7070 & $4\times900$ &   2.1 & 152 \\
         & 10.12.2015 & FPI       & \Ha        & $30\times150$&   1.2 &  \\
J1237+39 & 10.12.2016 & Long-slit & 3600--7070  & $4\times900$ &  1.0 & 333 \\
         & 12.12.2015 & FPI       & \Ha         & $30\times150$&   1.8 & \\
\hline
\end{tabular}
\end{table*}

\citet{Wong2015} has posited that these galaxies present different stages of quick star formation quenching driven by a kinetic process, possibly feedback from the black-hole activity.
According to the proposed evolution models, two of them (J0836+30 and J0900+46) appear to evolve along the fastest quenching evolutionary pathway, on timescales of up to 1 Gyr, typical of early-types; two others (J1117+51 and J1237+39) evolved less and favour slower quenching pathways, as the late-types do. 
\citet{Wong2015} suggested that  the start quenching time in these galaxies and fiducial quenching starting point in the models differ. However, the other scenario could be proposed:  J1117+51 and J1237+39 collected external gas from the outer source after quenching began and moved in the opposite direction in the colour-magnitude diagram.

However, the angular resolution of WSRT data (beam  12--30 arcsec)  was comparable with optical sizes of the galaxies; that prevents studying gas kinematics on the intergalactic scales. Spatially-resolved optical spectroscopy could help us understand what we observe in these galaxies: a time  sequence of AGN feedback \citep[as][proposed]{Wong2015}, or a more complex process including a new gas accretion event. In the present paper, we obtained    detailed observed data on kinematics and distribution of the ionized gas in galaxy discs in the \citet{Wong2015} sample and compared the gas (in both \HI\ and \HII\ species) and stellar kinematics.  Thus, we can make more detailed decisions about the processes that influence gas reservoirs and probably find some relations between \HI\  and \HII.

\section{Observations and data reduction}

Optical spectroscopic observations  of the \citet{Wong2015} sample were performed in December 2015 at the prime focus of the Special Astrophysical Observatory of the Russian Academy of Sciences (SAO RAS) 6-m telescope BTA with the  SCORPIO-2 multimode focal reducer \citep{SCORPIO2}. Observations consisted of two steps.  Each  galaxy was observed in the long-slit mode of the spectrograph in order to understand the character of the ionized gas large-scale distribution. Only in J1117+51 and J1237+39, the ionized gas emission lines were detected in their  discs. These two targets were mapped with the scanning Fabry--Perot interferometer (FPI) in the \Ha\ emission line. The   log of observations is presented in Table~\ref{tab:obs}.

\subsection{Long-slit spectroscopy}

\label{sec:obslong}

The  large-scale  distribution  of \HI\ in the sample galaxies  was taken into account to choose the slit position angles ($PA$). Namely, the slit was placed across the nucleus along the major axis of the \HI\ structures in the galaxies  J0836+30 and J0900+46, where \HI\ clouds according to \citet{Wong2015}  were misaligned with the stellar disc. In    J1117+51 and J1237+39 with the regular rotating \HI\ discs aligned with the stellar ones, the slit crossed  the nucleus along stellar disc major axis. 

Fig.~\ref{fig:sdss} shows the position of the SCORPIO-2 slit  with a size of 6~arcmin $\times$ 1~arcsec   in the SDSS images of the galaxies. The scale  was 0.36~arcsec per pixel, the total exposure time and typical mean are given  in Table~\ref{tab:obs}. Observations  were obtained with the VPHG1200@540 grism providing the spectral resolution $\delta\lambda \approx 5.0$~\AA.  Initial data reduction was performed in a standard way using the \textsc{IDL}-based software package written for reducing  the long-slit spectroscopic data obtained with \mbox{SCORPIO-2} as  described in our previous papers \citep[e.g.,][]{Egorov2018}. The flux calibration was done by using the spectrophotometric standard star observed at the same night in air-mass conditions close to those in the galaxy. 

\begin{figure*}
\centerline{\includegraphics[width=0.52\linewidth]{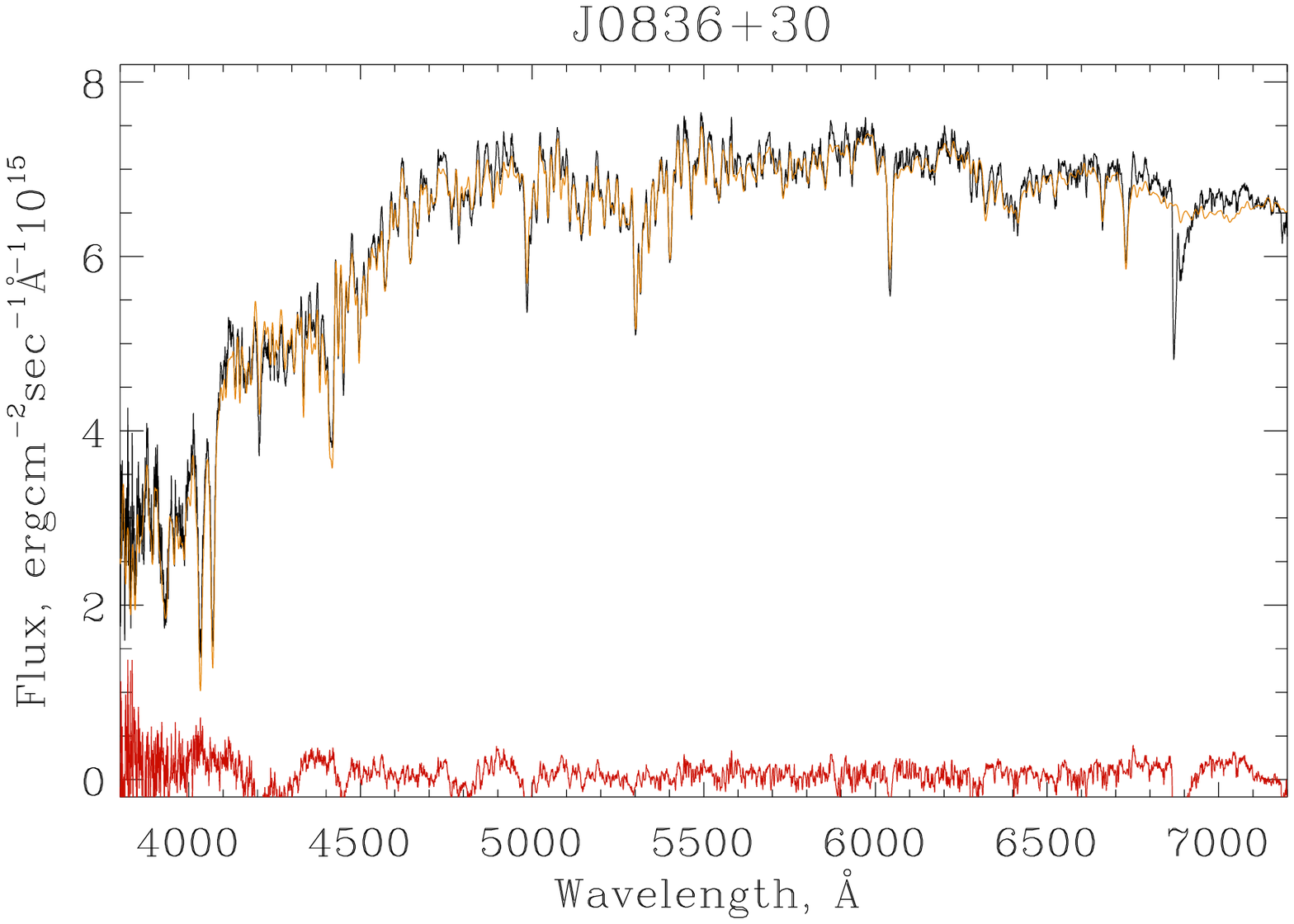}
\includegraphics[width=0.52\linewidth]{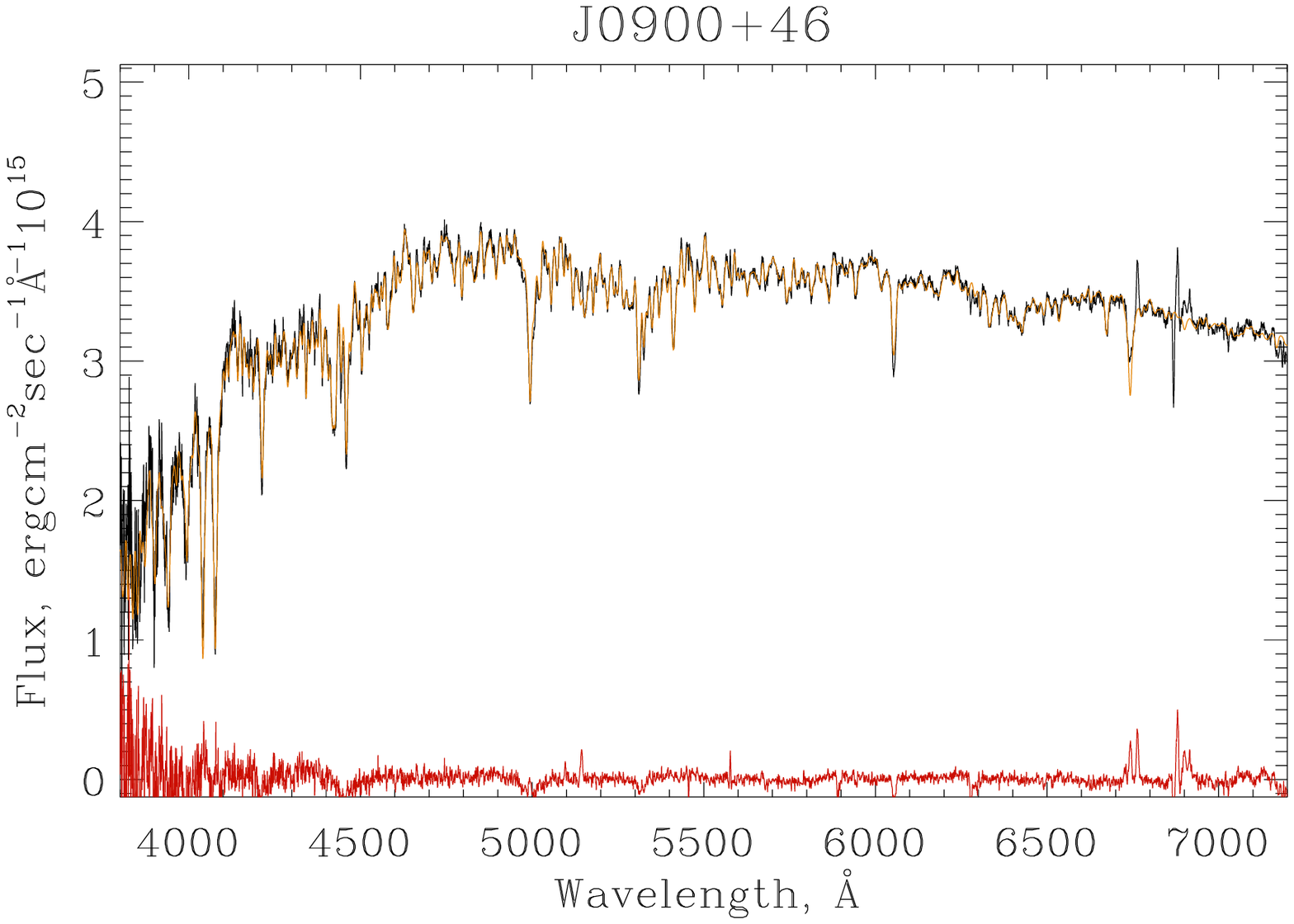}}
\centerline{\includegraphics[width=0.52\linewidth]{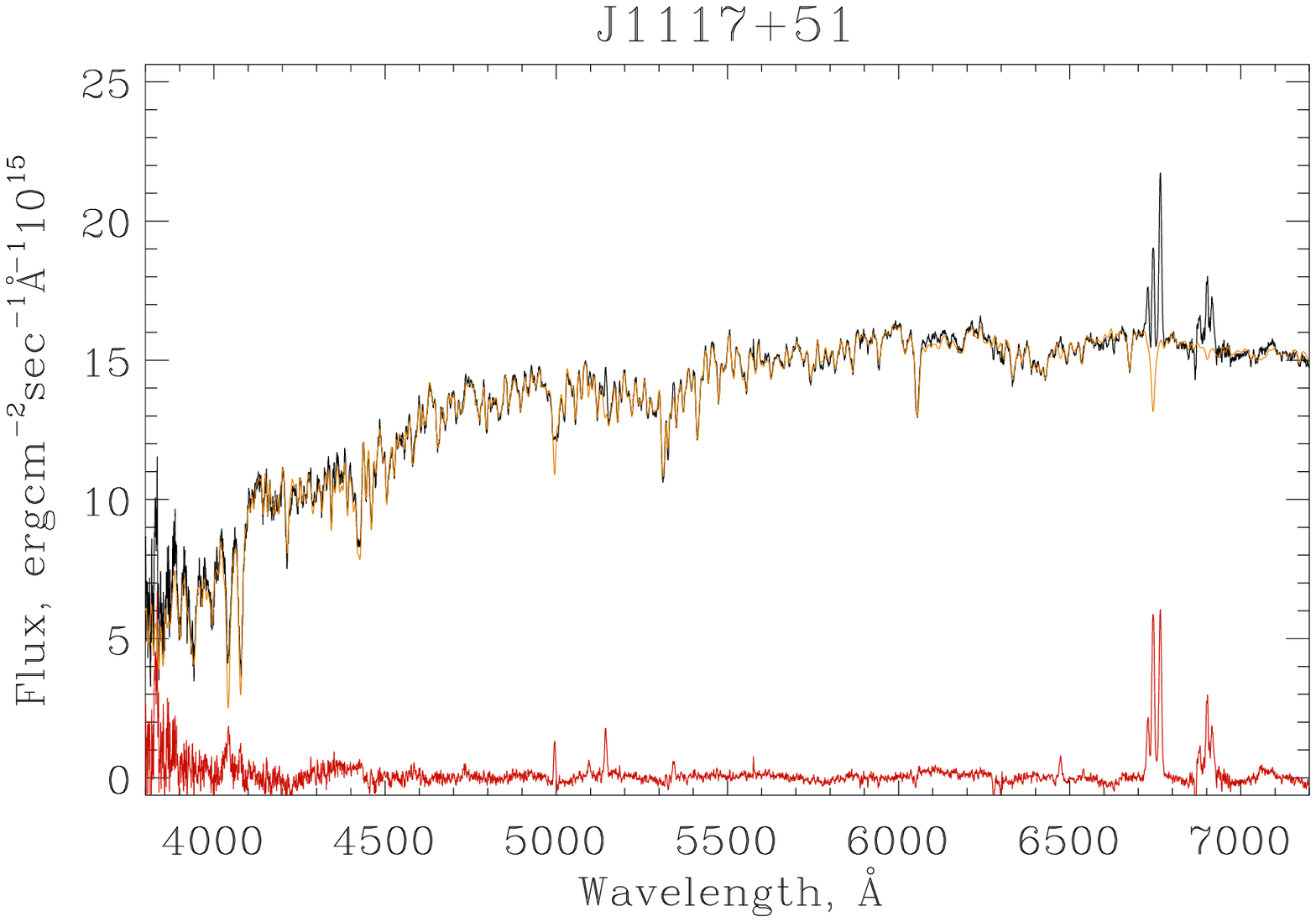}
\includegraphics[width=0.52\linewidth]{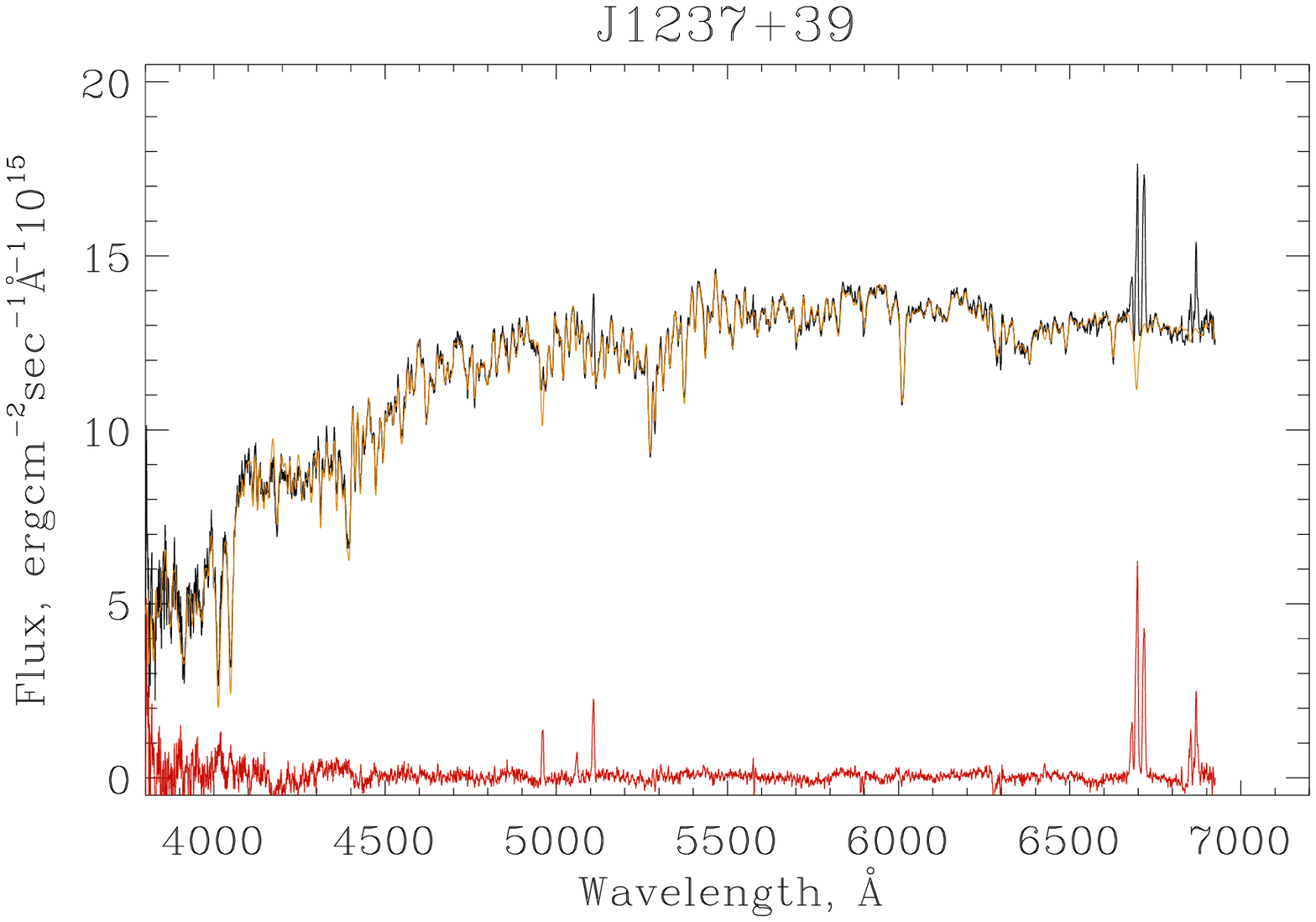}}
\caption{SCORPIO-2 spectra of the sample galaxies $\pm3$ arcsec integrated around the nucleus (black), the ULySS modelled spectra (yellow),  and the residual nebular spectrum (red). Top row: J0836+30 and J0900+46. Bottom  row:  J1117+51 and J1237+39. }
\label{nuc}
\end{figure*}

The observed long-slit spectra contain  the combination of the ionized gas emission lines with underlying stellar  absorption features. In order to analyse the nebular spectra, we subtracted the  stellar population contribution models using the
ULySS package  \citep{ULySS}. The Elodie and Vaz Miles stellar libraries were accepted. Fig.~\ref{nuc} shows the integrated  spectra of the galaxy nuclear  regions together with  its decomposition to the  stellar population and  ionized gas.
The emission lines were fitted by a single Gaussian (for \Ha\ and \Hb) or double Gaussian  (for doublets \SII$\lambda6717,6731$, \NII$\lambda6548,6583$, and \OIII$\lambda4959,5007$) to obtain their fluxes and line-of-sight velocities. 
In order to increase the signal-to-noise ratio, all the spectra used for further analysis were binned along the slit using a simple 2-3 px binning.

\subsection{3D spectroscopy with scanning FPI}

The extended  regions  of the ionized gas were detected in the SCORPIO-2 long-slit data in   J1237+39 and J1117+51. These two galaxies were also observed   with SCORPIO-2 in the scanning FPI mode. This device mapped the spectral region around the redshifted \Ha\  emission line in 40 spectral channels. The  corresponded  spectral range was about 37\AA, with a spectral resolution of 2.4\AA\ ($FWHM\approx110\km$). The data were reduced using an IDL-based software package \citep{MoiseevEgorov2008,Moiseev2015}. After the initial reduction, the observed data were combined into data cubes with a scale of 0.7 arcsec/px. The \Ha\ emission-line profiles were fitted  by the one-component Voigt function which yielded the flux, line-of-sight velocity, and velocity dispersion corrected for instrumental broadening \citep{MoiseevEgorov2008}.

\section{Ionized  gas in   J1117+51 and J1237+39 }

Fig.~\ref{nuc} clearly  demonstrates the absence of detectable emission lines in the nuclear spectra of J0836+30, also we did not see any diffuse nebular emission along the slit. The  similar situation was observed  in J0900+46, where only a weak nuclear emission was detected. In this   section, we consider the ionized gas properties only in  two galaxies with extended discs of the ionized gas: J1237+39 and J1117+51.

\begin{figure*}
\includegraphics[width=0.45\linewidth]{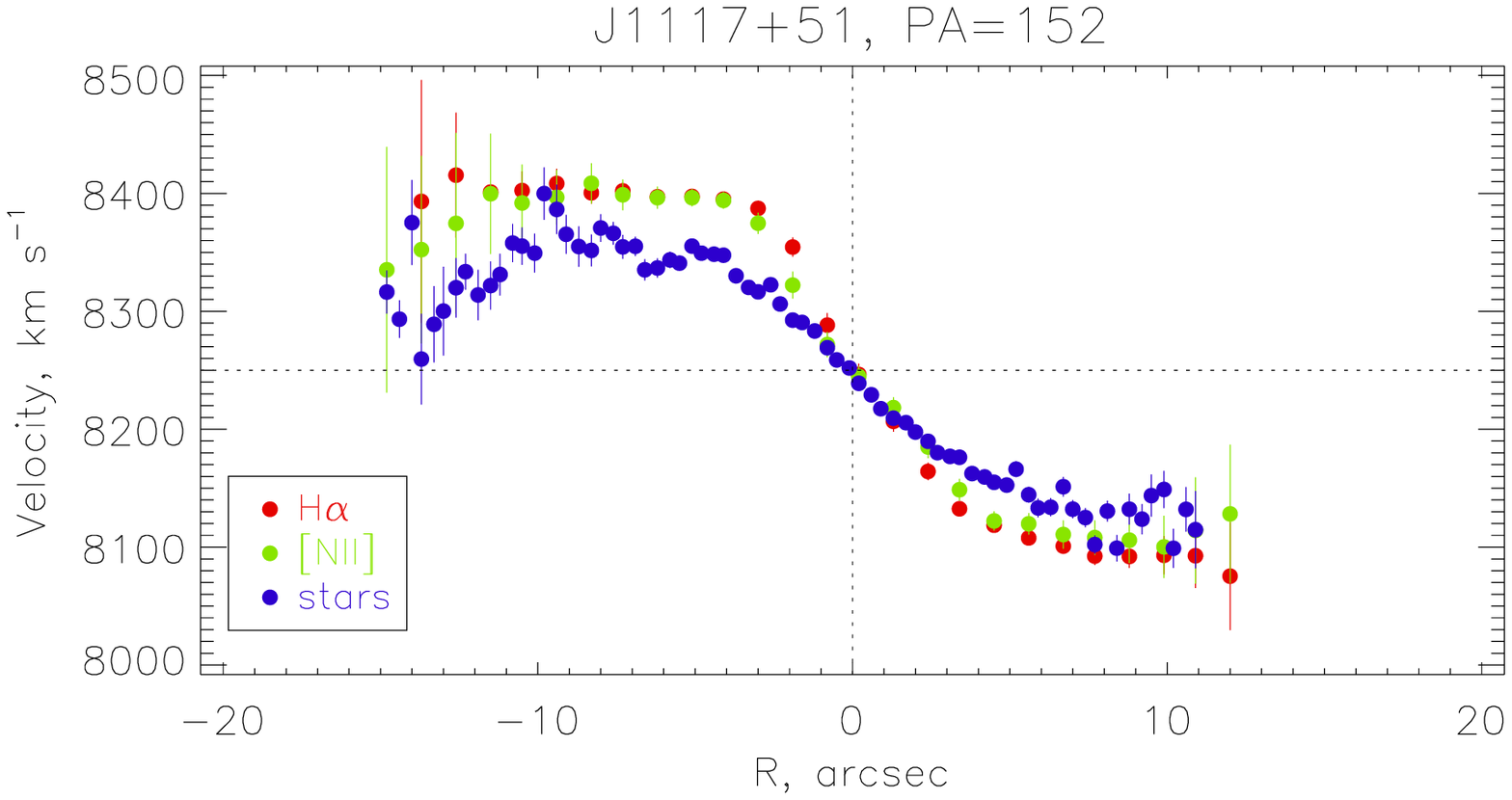}
\includegraphics[width=0.45\linewidth]{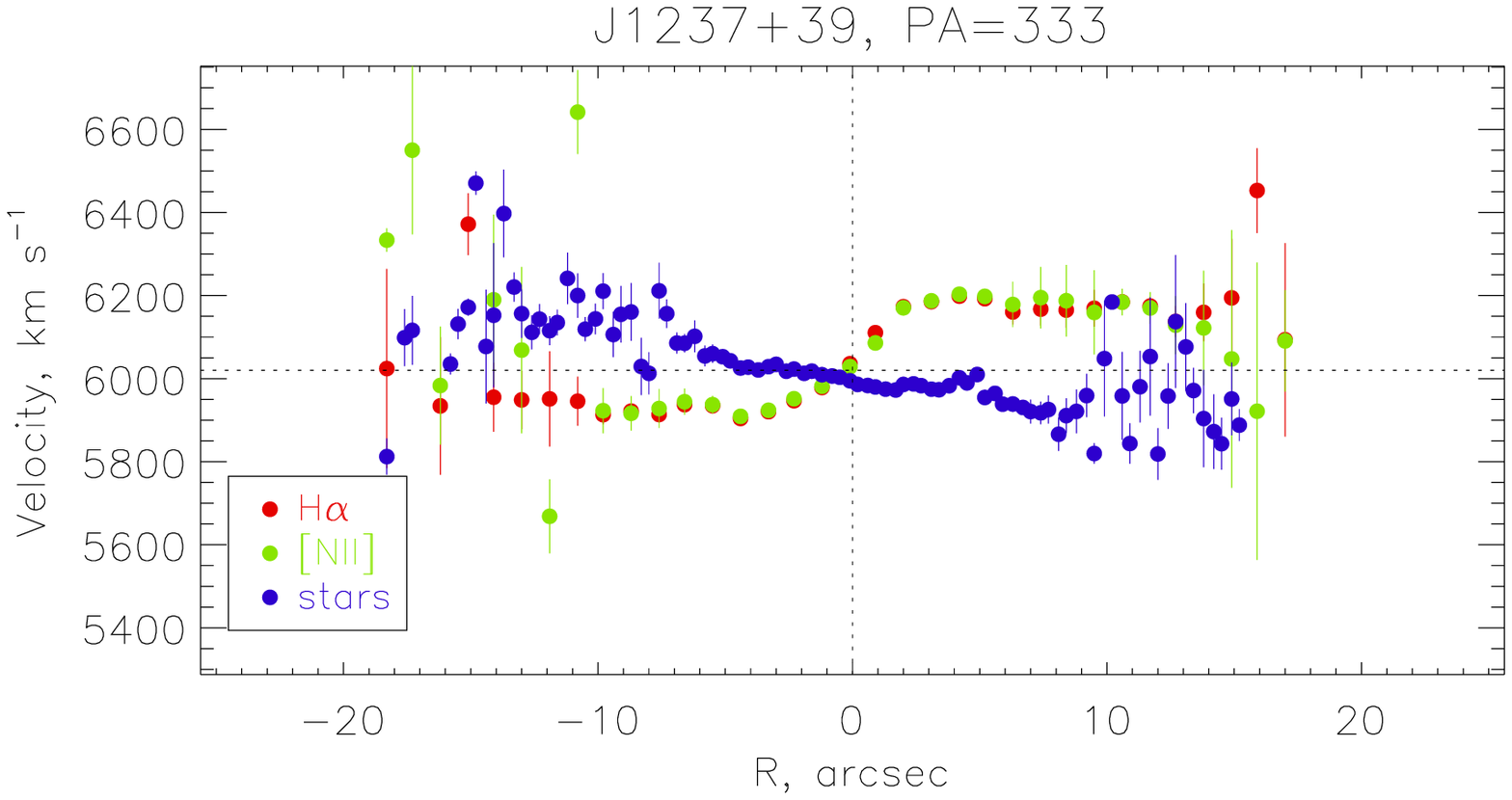}
\caption{Line-of-sight velocities of the stars (blue)  and of the ionized gas  in the  \Ha\ (red) and \NII$\lambda6548,6583$ emission lines (green) along the slits  in J1117+51 (left) and J1237+39 (right).
}
\label{lineofsight}
\end{figure*}

\subsection{Ionized gas kinematics  and morphology}

\label{sec:kin}
Radial distributions of the line-of-sight velocities of the stellar and gaseous components along the photometric major axes   in  J1117+51  and J1237+39   are shown in Fig.~\ref{lineofsight}.  The line-of-sight velocities of stars were derived by fitting the observed spectra with the ULySS  population model as described above in Sec.~\ref{sec:obslong}. 

\begin{figure*}
\centerline{
\includegraphics[height=0.3\linewidth]{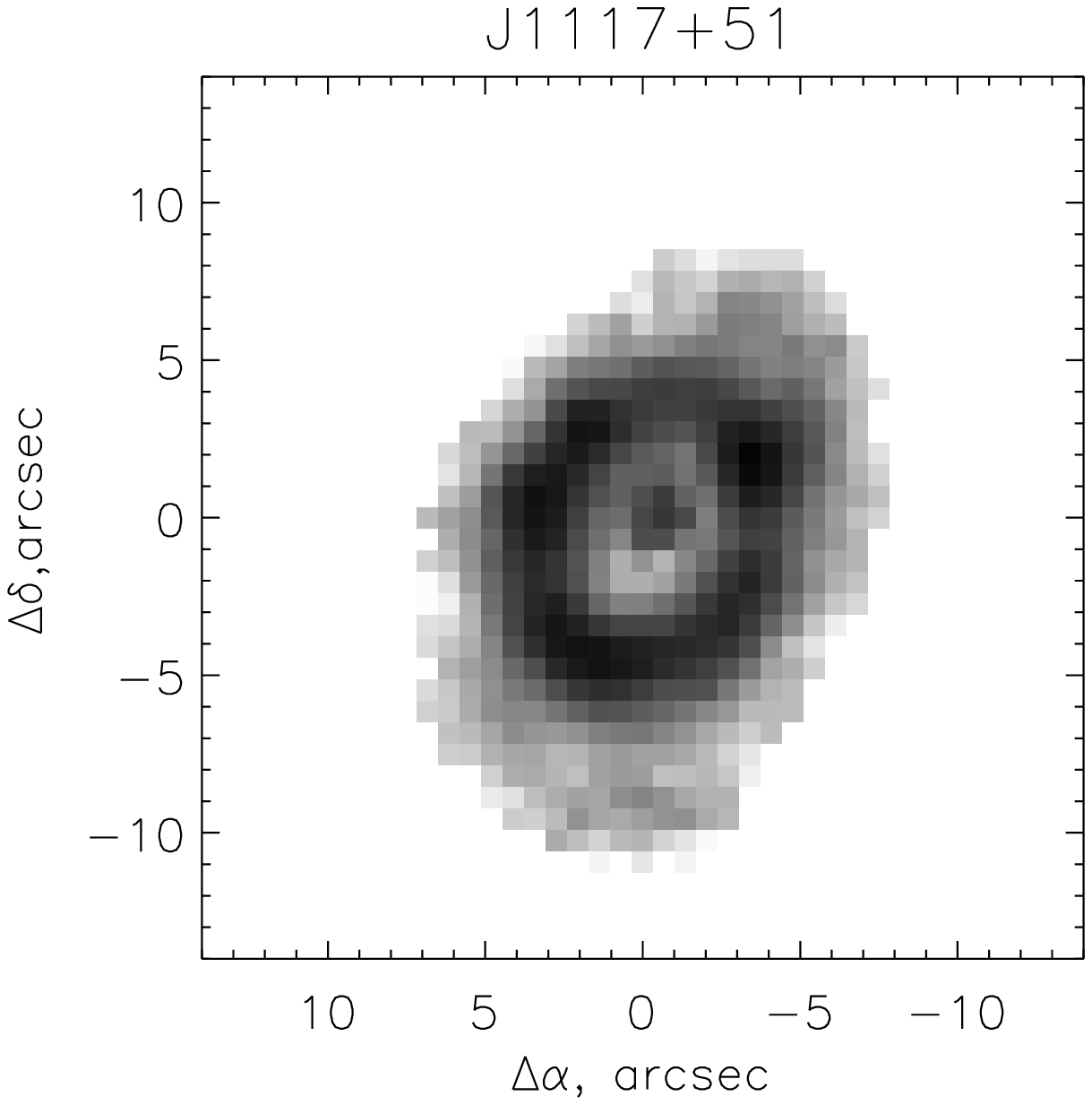}
\includegraphics[height=0.3\linewidth]{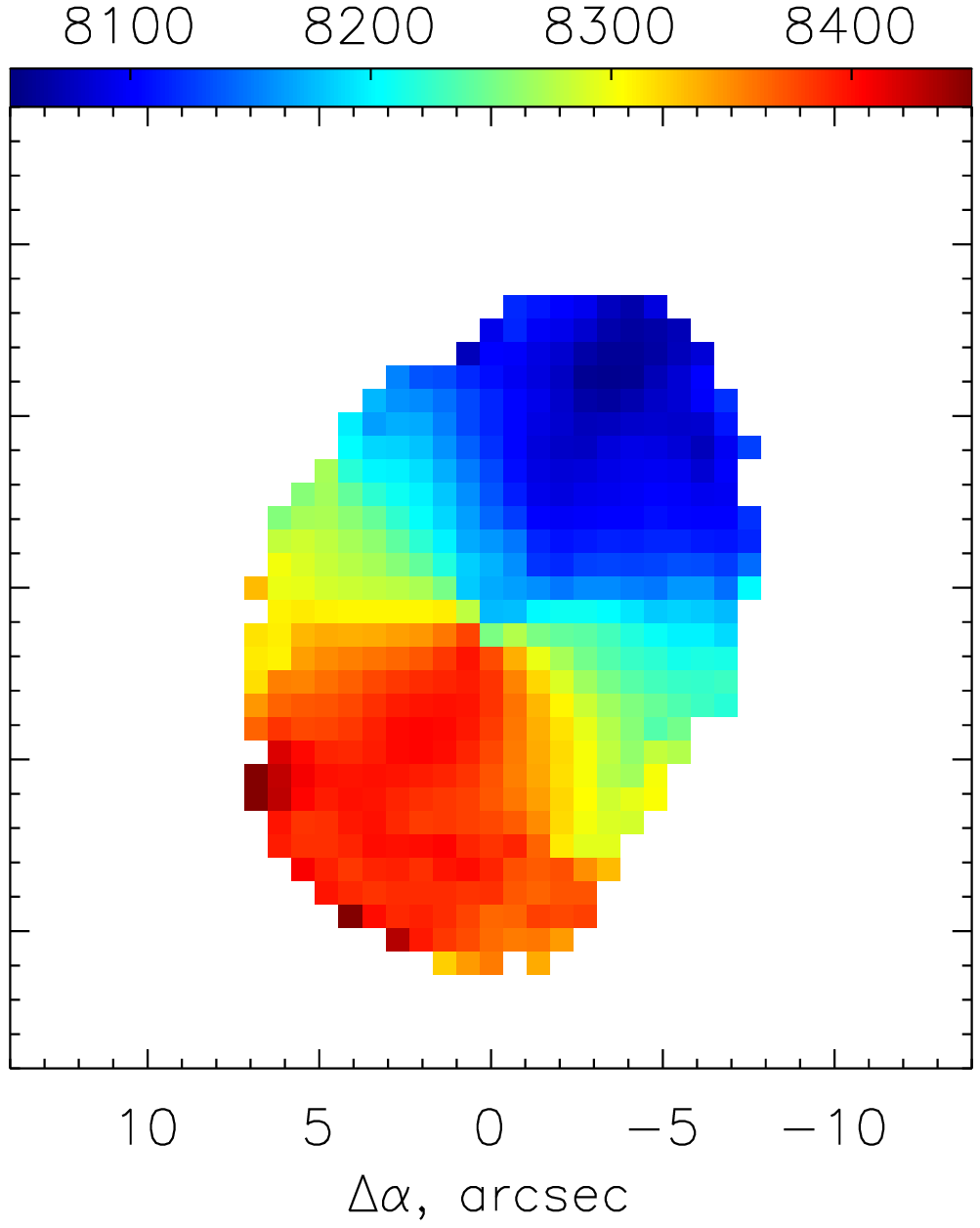}
\includegraphics[height=0.3\linewidth]{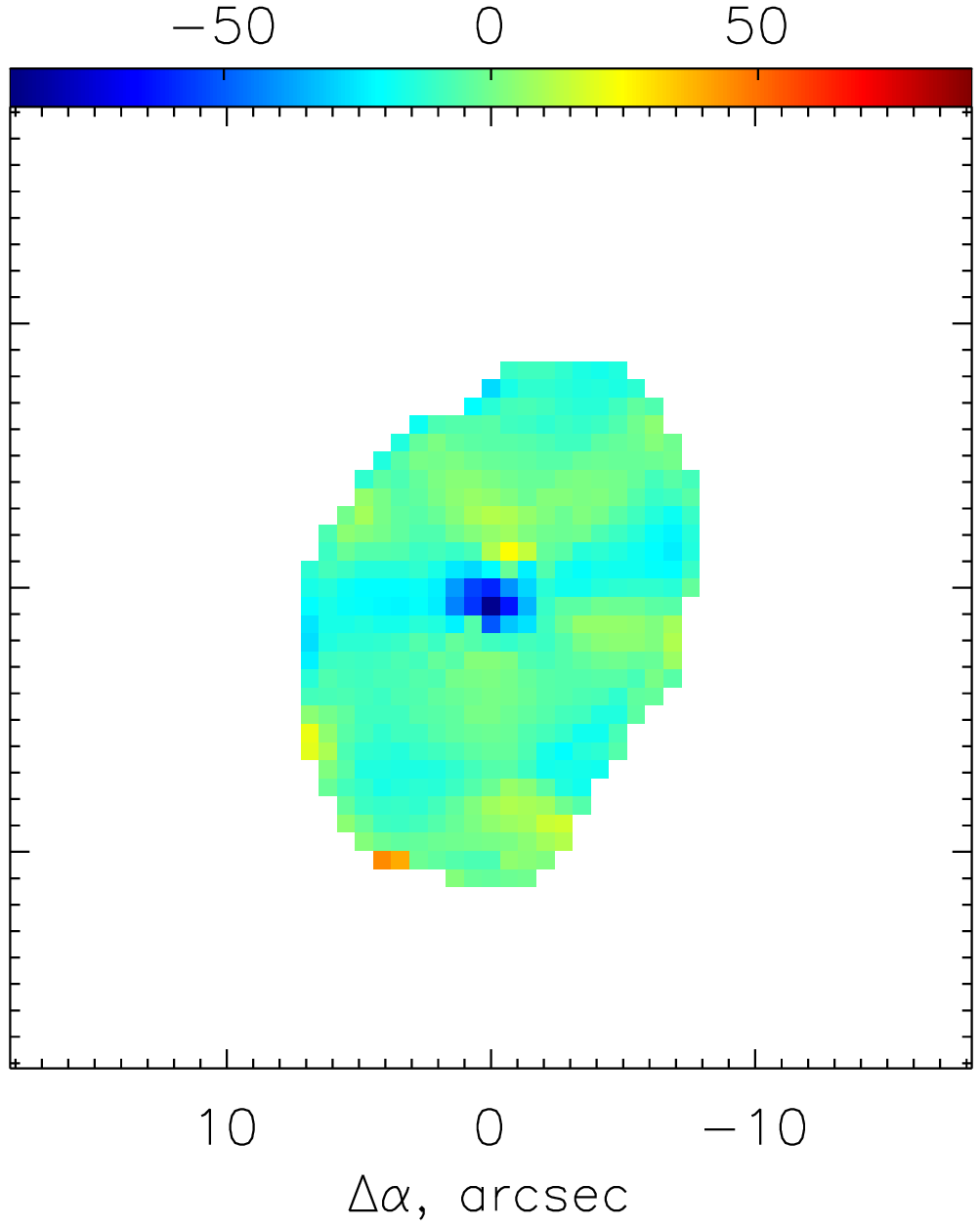}
\includegraphics[height=0.3\linewidth]{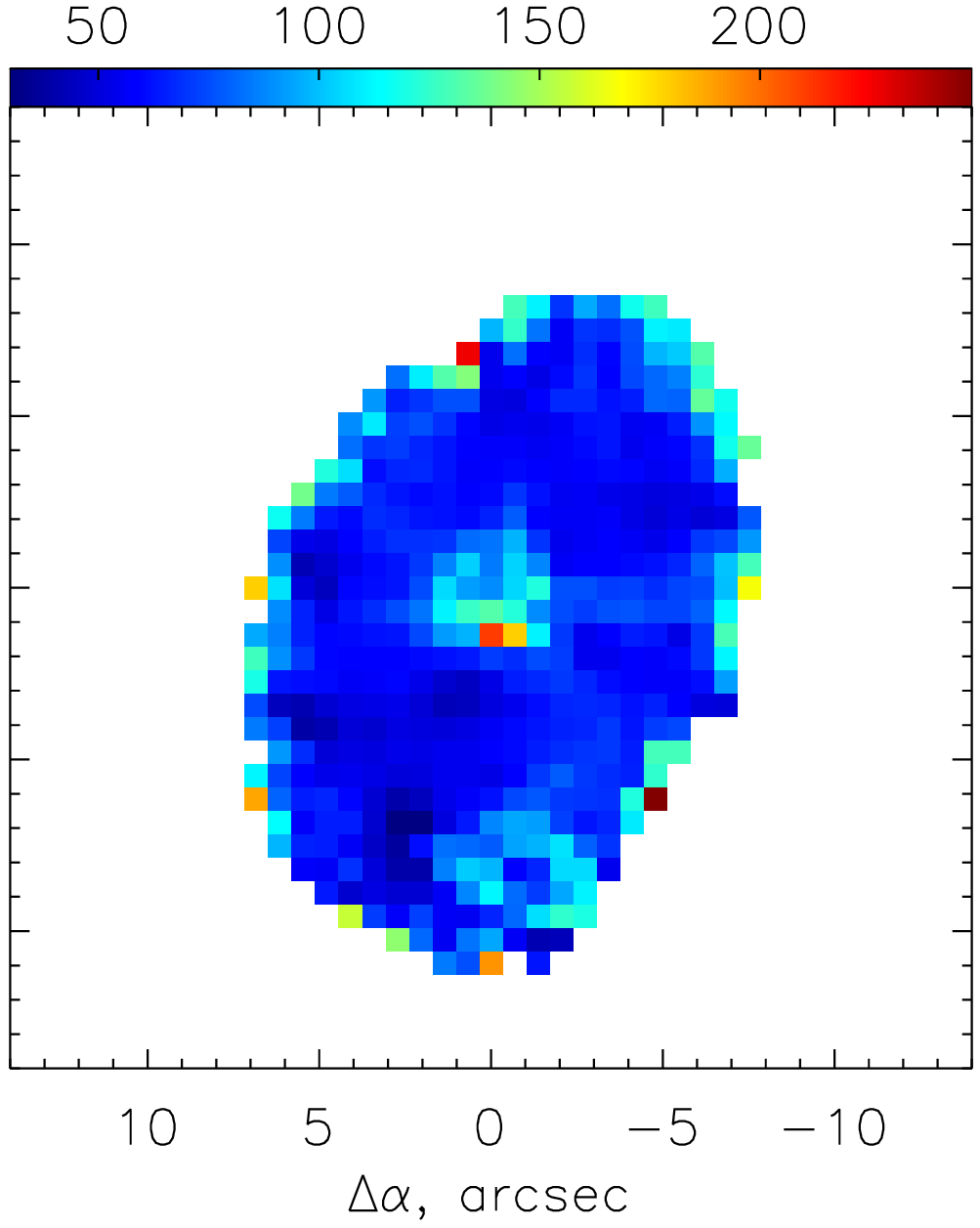}
}
\centerline{
\includegraphics[height=0.3\linewidth]{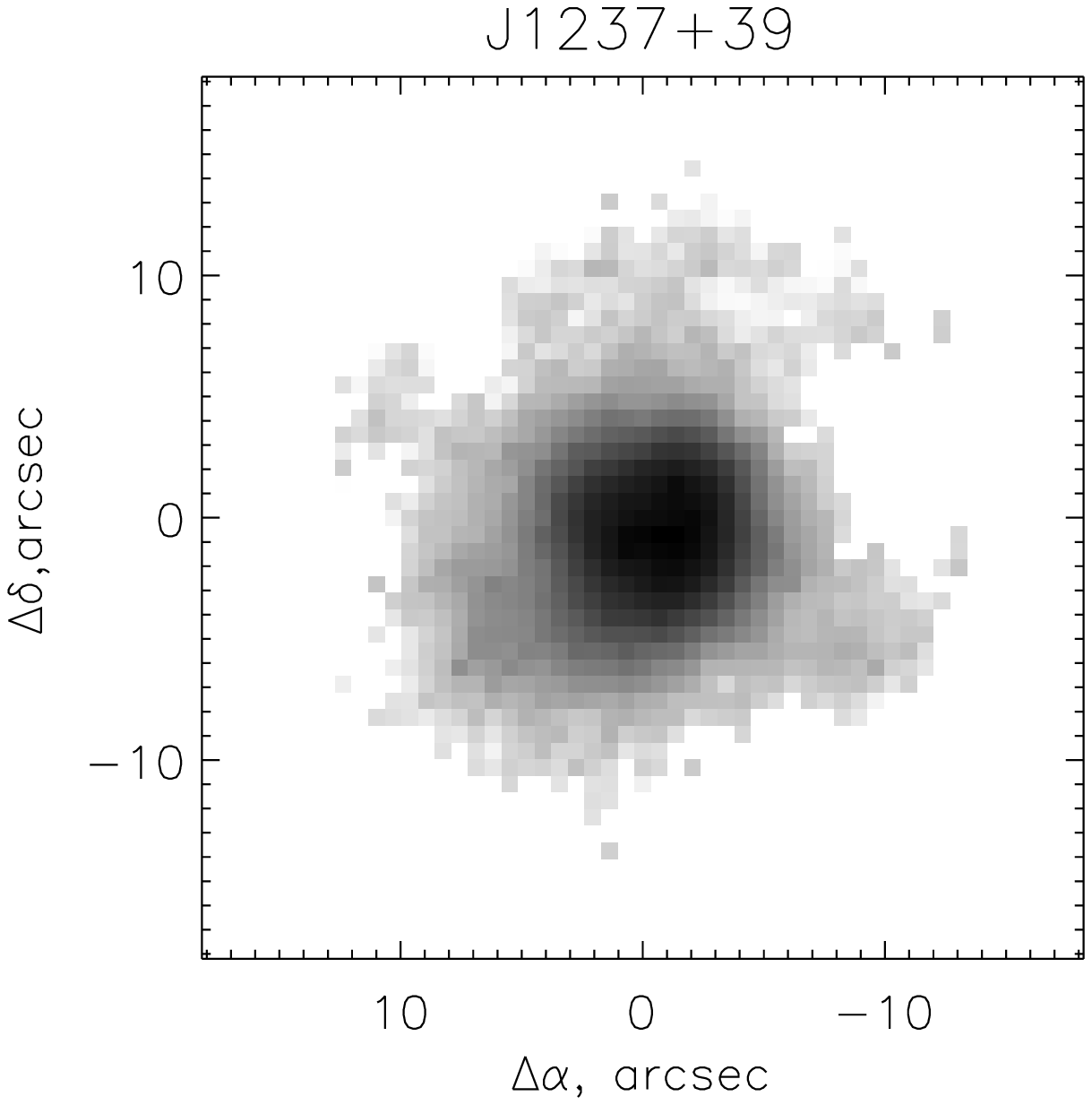}
\includegraphics[height=0.3\linewidth]{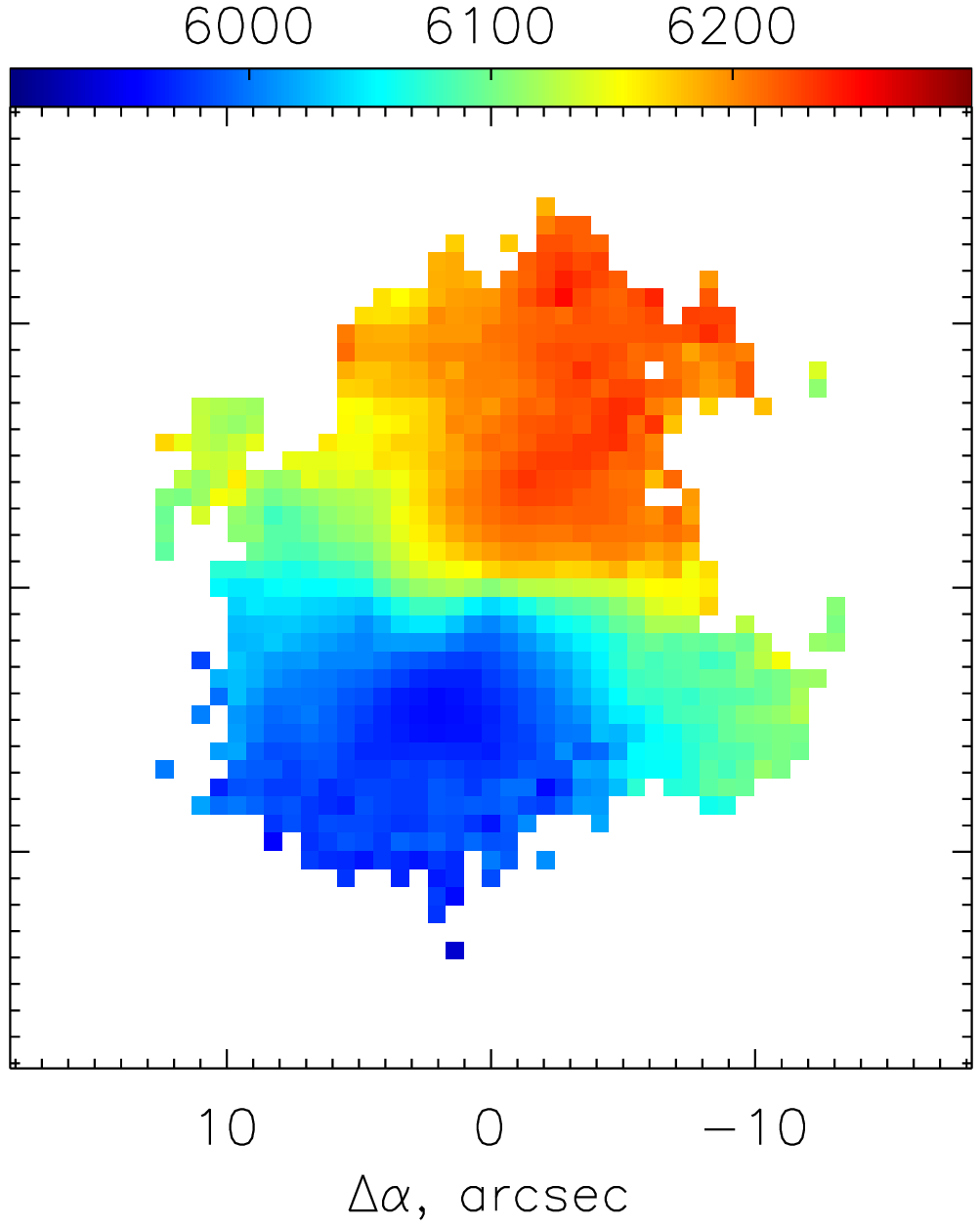}
\includegraphics[height=0.3\linewidth]{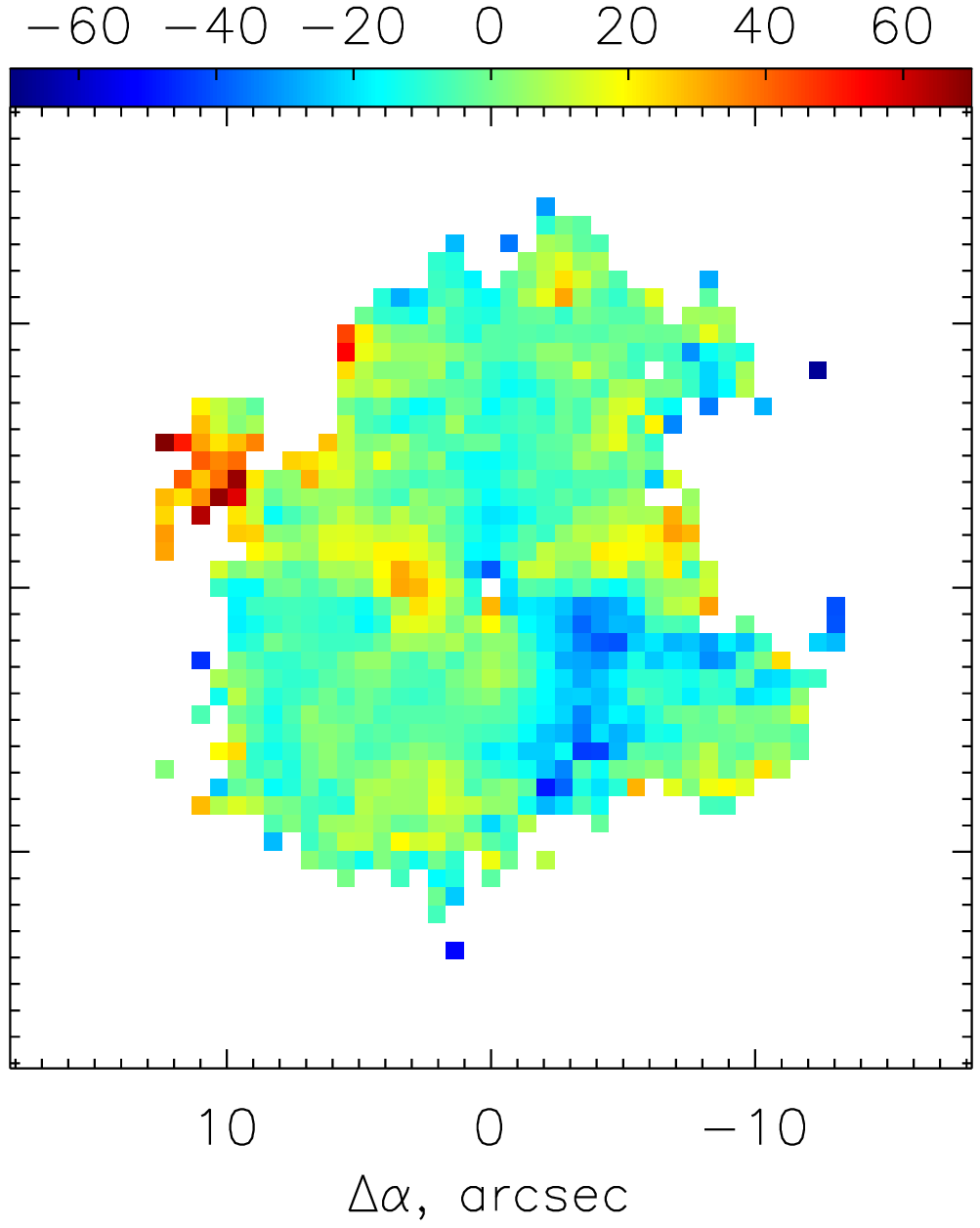}
\includegraphics[height=0.3\linewidth]{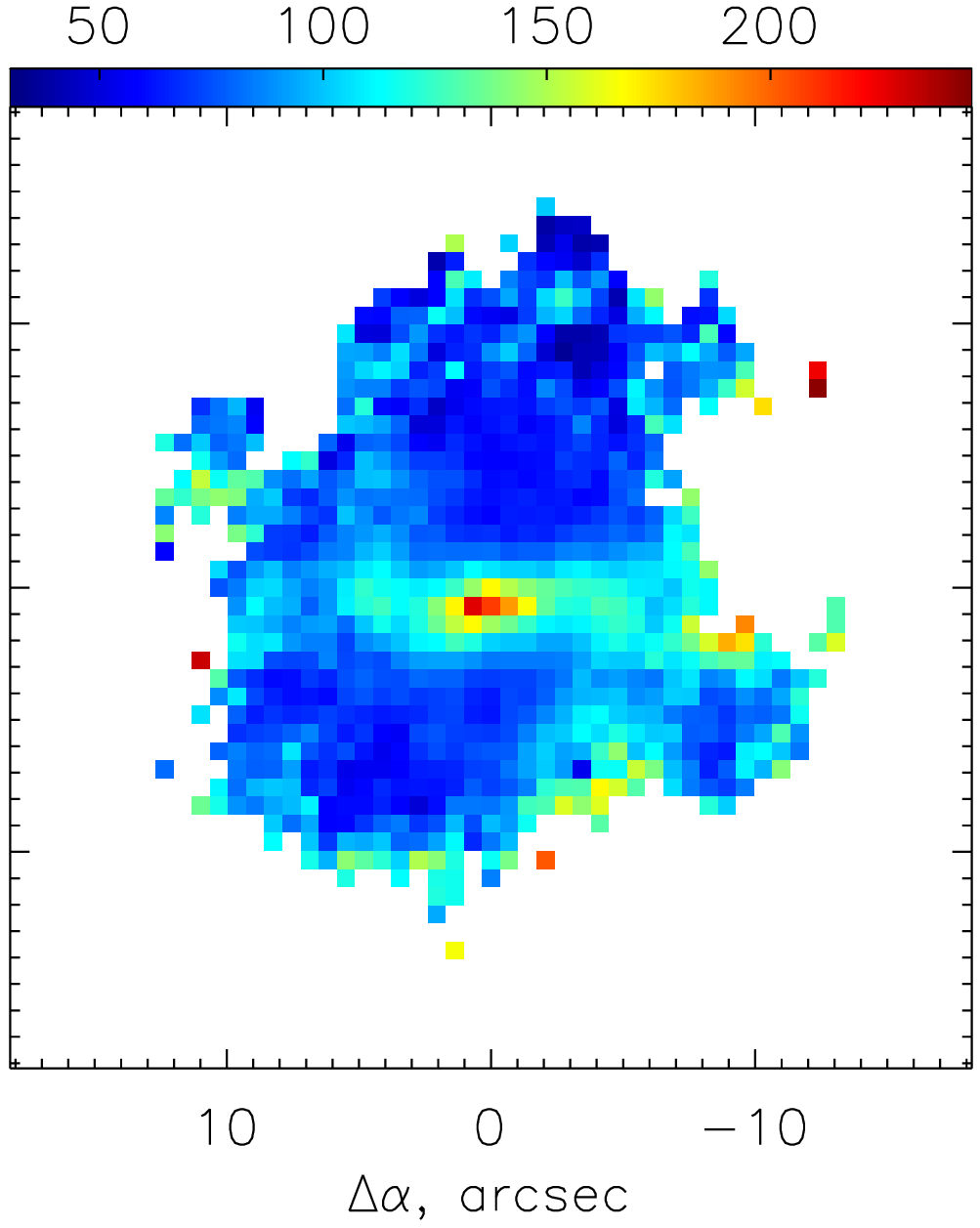}
}
\caption{FPI maps in the \Ha\ emission line for J1117+51 (top row) and J1237+39 (bottom row). From left to right: the monochromatic images in the logarithmic scale, the line-of-sight velocity field, the maps of residual velocities (observed minus tilted-ring model) and  velocity dispersion. Point $(0,0)$  corresponds to the nucleus in continuum, the colour scale is in $\km$.  
}
\label{fig_ifp}
\end{figure*}

In both galaxies, we see a typical kinematic  pattern  of a regular rotating disc: the   velocities of each component demonstrate symmetrical  distributions relative to the galaxy nucleus and corresponding systemic velocities; the observed amplitude of the dynamically cold gas is larger than that of stars (i.e., an asymmetric drift). The FPI \Ha\ velocity fields (Fig.~\ref{fig_ifp}) also confirm  the regular rotation pattern of the ionized gas disc with the similar line-of-sight  velocity changes that have been observed  in  \HI. Namely, \citet{Wong2015}  have shown that SE side of the \HI\ disc in J1117+51 is redshifted relative to the systemic velocity in agreement with our   data in \Ha. Whereas in the FPI maps the circularly rotating disc is resolved in details thanks to the angular resolution in \Ha\ one order higher than that in \HI. J1237+39 is the only galaxy with the well-resolved rotating \HI\ disc in the \citet{Wong2015} paper (see their Fig.~7b), the neutral gas velocity field seems to be in a very good agreement with the FPI data, including even small-scale details like an extension of `green' velocities along the disc minor axis.  

The principal difference between the considered galaxies is that in J1117+51 the gas and stars  rotate in the same direction, whereas in J1237+39 all the observed gas (\HI+\HII) is in the counter-rotation. The diameter of the gaseous disc is about 28\arcsec (=11 kpc) in \Ha\ and it exceeds 30 kpc in \HI, that is even larger than the stellar disc diameter (D$_{25}$=21 kpc in the r-SDSS according to NED). Such global large-scale gaseous counter-rotating disc is known in several nearby galaxies \citep[][and references therein]{Silchenko2009,Pizzella2018}. Recently, a few dozen similar gas-star misaligned structures were found in the 1st release of the MaNGA survey \citep{Jin2016}.

\begin{figure*}
\centerline{
 \includegraphics[width=0.5\linewidth]{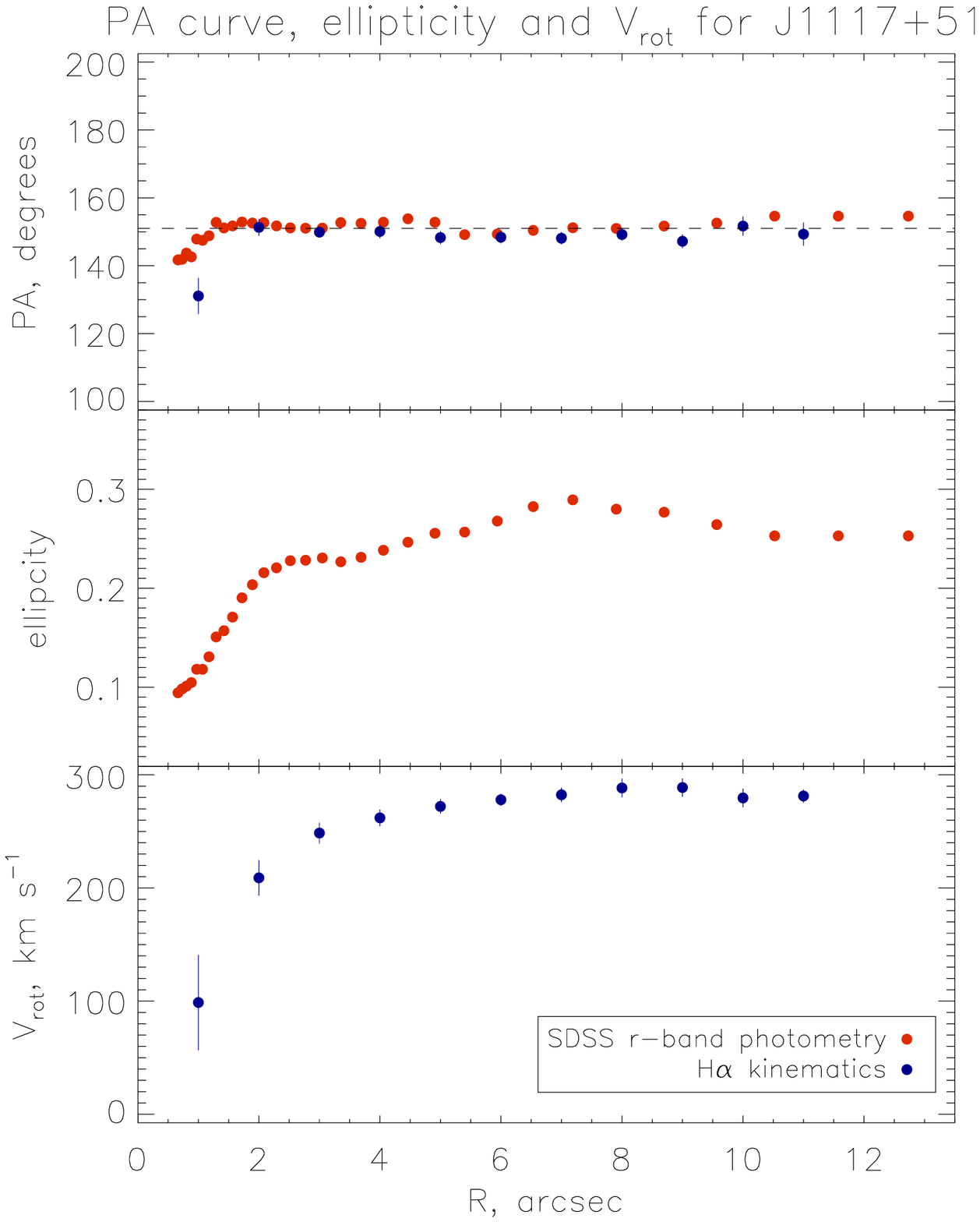}
 \includegraphics[width=0.5\linewidth]{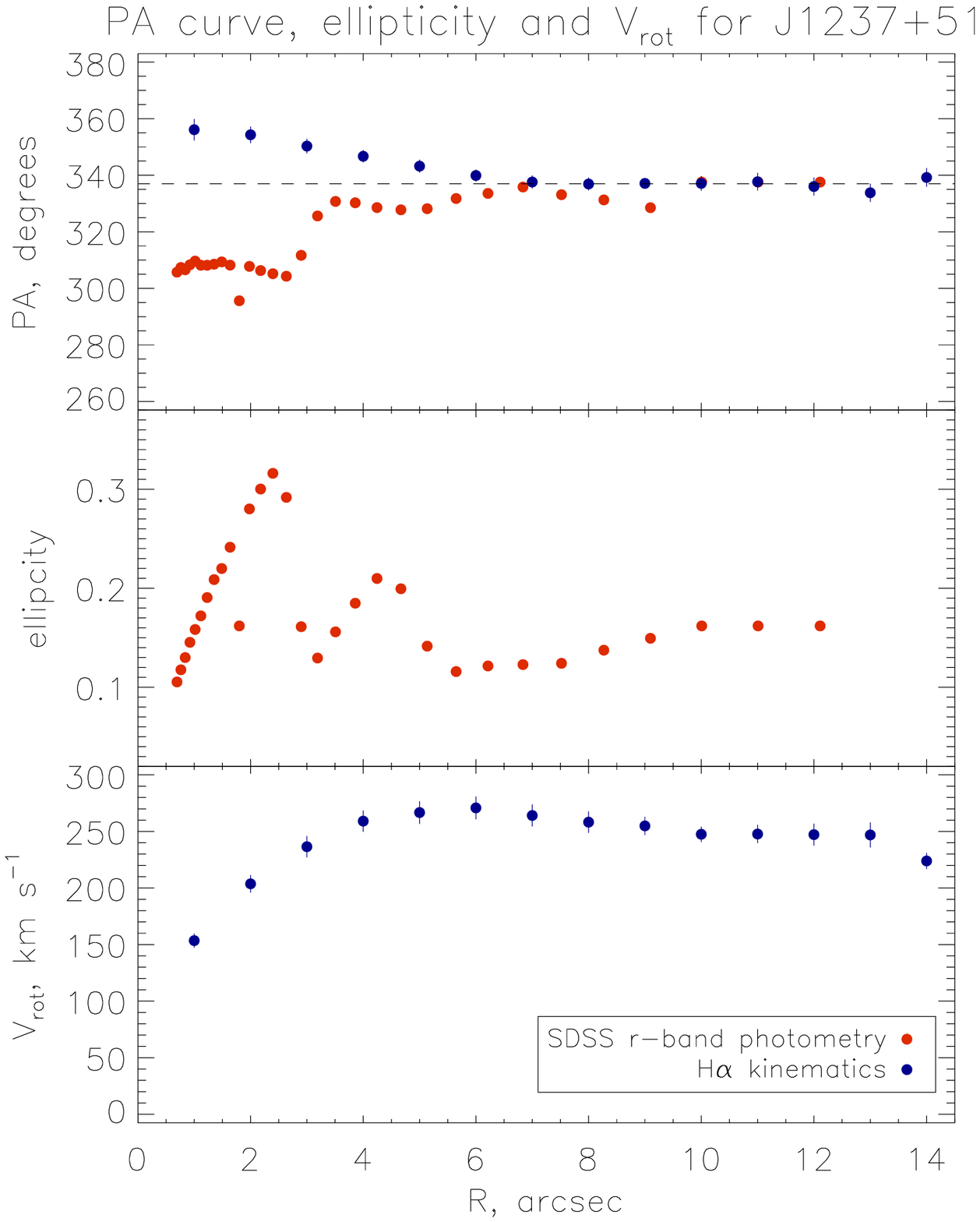}}
\caption{Radial variations of the tilted-ring model and $r$-band isophote parameters: the position angle (top), the ellipcity (middle), and the circular rotation velocity (bottom). The $PA$ of isophotes in r-band SDSS image is shown with red dots,  \HII\ kinematics -- blue dots. Left-hand panels --  J1117+51, right-hand panels -- J1237+39. The dashed lines correspond to the $PA_{0kin}$ line-of-nodes disc from Table \ref{tab:discs}.}
\label{fig:PA}
\end{figure*}

The \Ha\  velocity fields  were analysed with the classical `tilted-ring'   model \citep{Begeman1989} using the technique    described   in \citet{Finkelman2011,Moiseev2014}. The velocity fields  were splitted into  elliptical rings in agreement with the adopted inclination ($i_0$) and the position angle of the disc major axis ($PA_0$). The kinematic center in both galaxies matches the position of the photometric nucleus very well. In each ring, the observed distribution of the line-of-sight velocities was described with the circular rotation model parameterized by the position angle of the kinematic axis $PA_{kin}$, the inclination of circular orbits $i$, the rotation velocity $V_{rot}$, and systemic velocity $V_{sys}$.  The preliminary estimates of  $i_0$ and  $PA_0$ were derived from the analysis of the $r$-SDSS isophotes using the PhotUtils Python package\footnote{https://zenodo.org/record/2533376}.  $PA_0$ was clarified in the forthcoming tilted-ring analysis. Since the  kinematic value $i$ was estimated with large uncertainty,  we kept its photometric values obtained from the $r$-band external isophotes using the intrinsic axial ratio $q_0=0.2$ adopted for spiral galaxies \citep{Hubble1926}.   The mean values of  $V_{sys}$ were fixed after adopting the optimum value. The obtained parameters of the gaseous discs are listed in Table~\ref{tab:discs} including the orientation parameters obtained from the analysis of outer isophotes ($r>10$ arcsec). The radial dependences of $PA_{kin}$ and $V_{rot}$ are shown in Fig.~\ref{fig:PA}.

\begin{table}
	\centering
	\caption{Derived properties of the gaseous discs.}
	\label{tab:discs}
\begin{tabular}{l|r|r} \hline  
Parameter           &  J1117+51 & J1237+39  \\
\hline
$i_0$, $\deg$ 		&  $43\pm 1$  & $34\pm1$  \\
$PA_{0kin}$, $\deg$	& $151\pm 6$   &  $337\pm2$ \\
$PA_{0phot}$, $\deg$	& $152\pm 2$   &  $331\pm6$ \\
$V_{sys}$, $\km$      & $8240\pm 7$& $6095\pm5$	\\
\hline 
	\end{tabular}
\end{table}

Both velocity  fields are well described by domination of pure  circular gas motions with the flat rotation curve (Fig.~\ref{fig:PA}, bottom). To quantify the value of gas-star misalignment we compare photometric and kinematic estimations of the major axis $PA$. The top panels in Fig.~\ref{fig:PA}  as well as Table~\ref{tab:discs} clearly demonstrate  a good agreement between the $r$-band isophotes position angle ($PA_{phot}$)    and gaseous $PA_{kin}$ in the outer parts of the discs. However, the situation within the  central kiloparsecs is more complex.

Possible non-circular motions in the disc of J1117+51 are negligible, because the radial variations of $PA_{kin}$ close to the mean value are underestimation errors. Relatively small changes of  $PA_{phot}$  at  $r<7$ arcsec (4 kpc) might be related to the star-forming ring that has appeared in the \Ha-image (Fig.~\ref{fig_ifp}) and is also visible in the SDSS image  (Fig.~\ref{fig:sdss}).  
The internal ellipticity of this ring  caused the observed variations  of isophotes shape, but  it doesn't affect the velocity field.    

The map of residual velocities after subtraction of the tilted-ring model from the observed velocity field in  J1117+51 reveals the excess of negative velocities ($V_{res}=-90\km$) in the nucleus of J1117+51. The most obvious interpretation of this blueshifted gas is an AGN-driven outflow which is often observed in  Seyfert/LINER  galaxies 

\begin{figure}
\includegraphics[width=\linewidth]{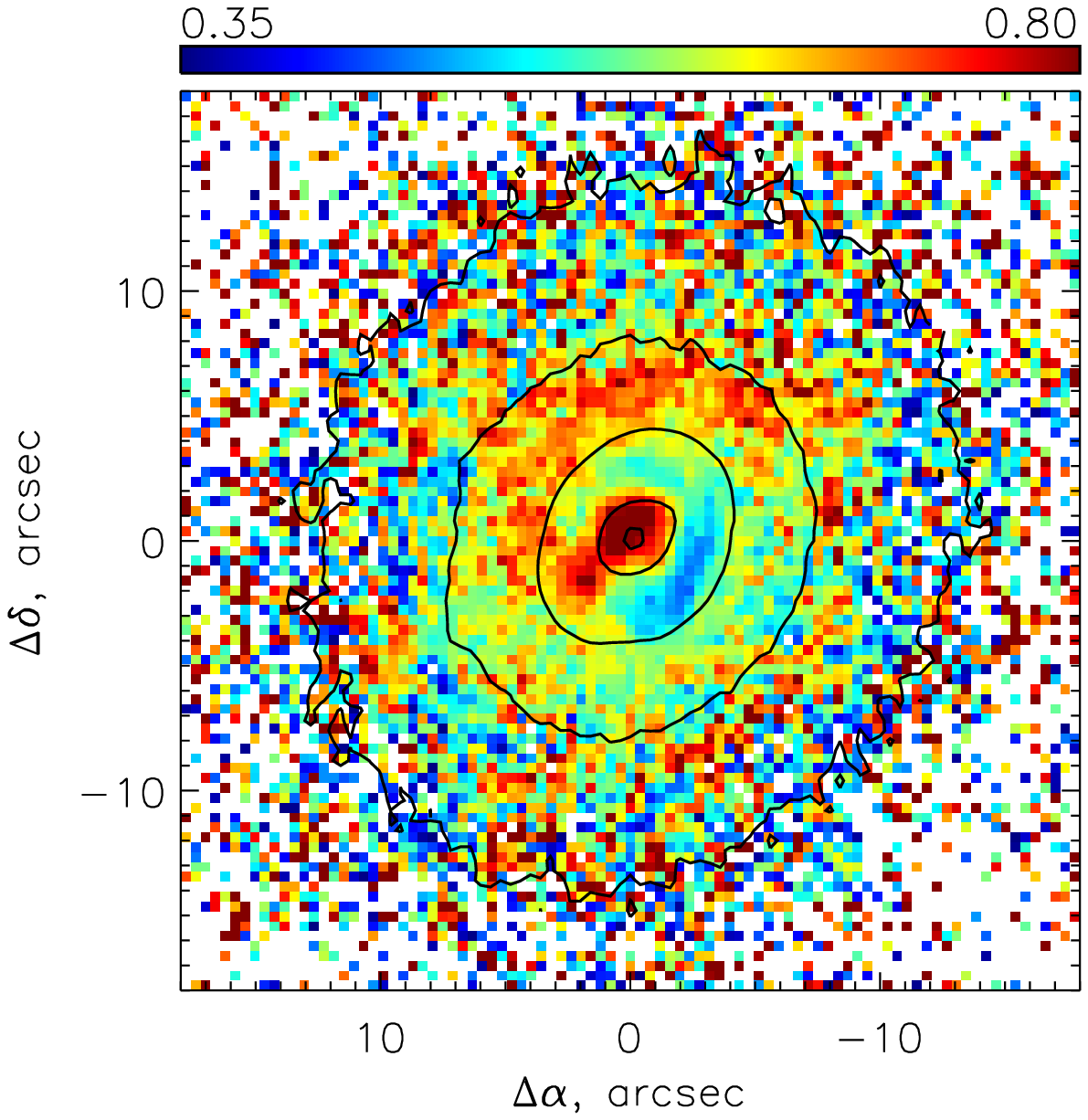}
\caption{\bf The SDSS map of the $(g-r)$ colour index in the galaxy J1237+39. Isophotes of the $r$-SDSS image are overlapped.}
\label{fig:sdss1237}
\end{figure}

The deviation from a simple rotating flat disc model  is more significant  in  J1237+39: the turn of  $PA_{kin}$ by 10--15 deg relative to the disc major axis $PA_0$ is observed in the central $5$ arcsec (2 kpc). In the same radii, $PA_{phot}$ deviates in the opposite direction (Fig.~\ref{fig:PA}). This sort of kinematic mismatch usually  corresponds to gas streaming motions under gravitational influence of a bar or a  triaxial bulge;  for detailed description and references see, for example, \citet{Moiseev2004}. Stationary values of the inner isophotes $PA$ are accompanied by the peak of ellipticity that is typical for bar-like structures \citep{Erwin1999}. The SDSS image (Fig.~\ref{fig:sdss}) reveals a low-contrast feature like an oval or pseudo-ring  at these distances, but the detailed analysis demonstrates a more complicated picture. Indeed, the SDSS colour map of $(g-r)$ shown in Fig.~\ref{fig:sdss1237} reveals several interesting details: a blue ring or pseudo-ring with $(g-r)\approx0.45-0.55$ at  $r=2-4$ arcsec (0.6--1.6 kpc) and a one-armed red spiral arm  $(g-r)\approx0.65-0.75$ extending to $r\approx8$ arcsec (3.3 kpc). This spiral starts from the red knot ( $(g-r)\approx0.72-0.81$) located at 1.2 kpc southeast from the galaxy nucleus. Very preliminary we can interpret the observed as minor merging imprints. In this case, the red knot may be a remnant of a dwarf companion galaxy, the red spiral corresponds to the stars of this disturbed satellite, or to a tidal structure generated by  merging, whereas a blue pseudo-ring is related to the star-formation burst  triggered by the concentration of the gas exchanging its angular momentum due the merging and falling in the cirucmnuclear region.

Therefore, we can conclude that the gas velocity field  in  J1237+39 is  perturbed by  asymmetric stellar structures  in the inner 3 kpc related to the minor merging remnants which produce the significant ($~30$ degs) deviations of kinematics $PA$  from the orientation of the isophotes   (Fig.~\ref{fig_ifp}).

The map of residual velocities in J1237+39 reveals several regions with significant ($\pm40$--$60\km$)  deviations from the tilted-ring model mostly along the galaxy minor axis (Fig.~\ref{fig_ifp}). It is not easy to describe  them in terms of an AGN-driven outflow or jet-cloud interaction, because the  gas around  the nucleus  has  a relatively low excitation, no extended narrow-line regions have been detected (see Sect.~\ref{sec:BPT}). Also, \citet{Wong2015} did not detect any elongation of non-thermal radio emission in this source. We suppose that  the observed  non-circular motions together with the extension of the diffuse \Ha\ emission along the galaxy minor axis \citep[that is also observed in \HI, see Fig.~7b in][]{Wong2015} might be related to the event formed in the global counter-rotating disc, either by gas accretion or minor merging whose possible remnants are detected in the distribution of the $(g-r)$ colour index.

\subsection{Gas excitation}
\label{sec:BPT}

\begin{figure*}
    \includegraphics{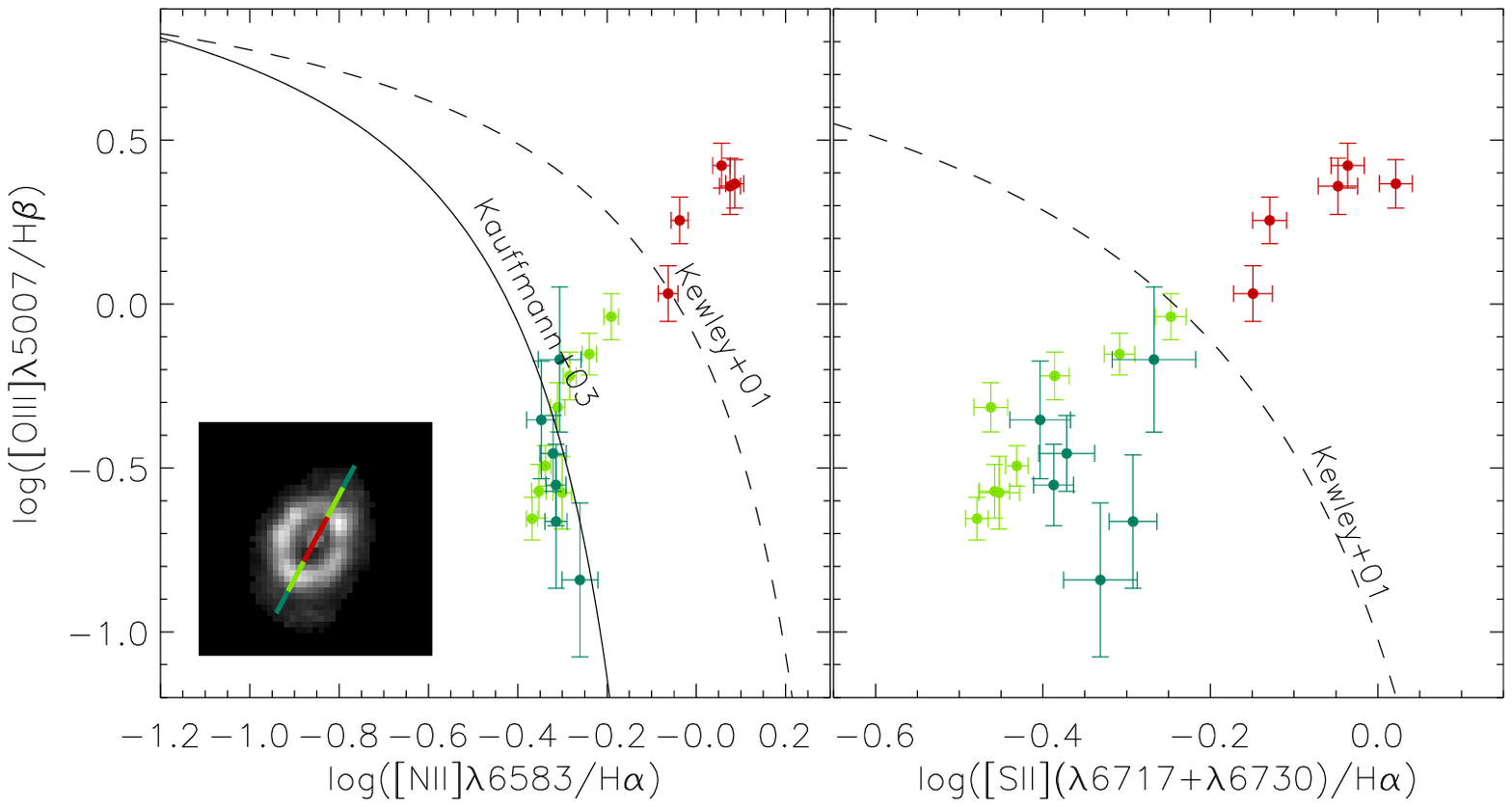}
    \caption{J1117+51 diagnostic diagrams of \OIIIHb\ versus  \NIIHa\  (left-hand panel) and \SIIHa\ (right-hand panel). Lines of separating pure \HII\ regions from AGN/shocks and composite excitation mechanisms: the dashed line is the \citep{Kewley2001} curve, the solid line is the \citep{Kauffmann2003} curve.  Different colours correspond to the different regions according to the sketch in the inset (\Ha\ image) with the overlapped SCORPIO-2 slit:  the nucleus (red), the star-forming ring (light green), and the external disc (dark green). Error-bars correspond to the 1$\sigma$ error.
    }
   \label{fig:bpt1117}
\end{figure*}

\begin{figure*}
\includegraphics[width=\linewidth]{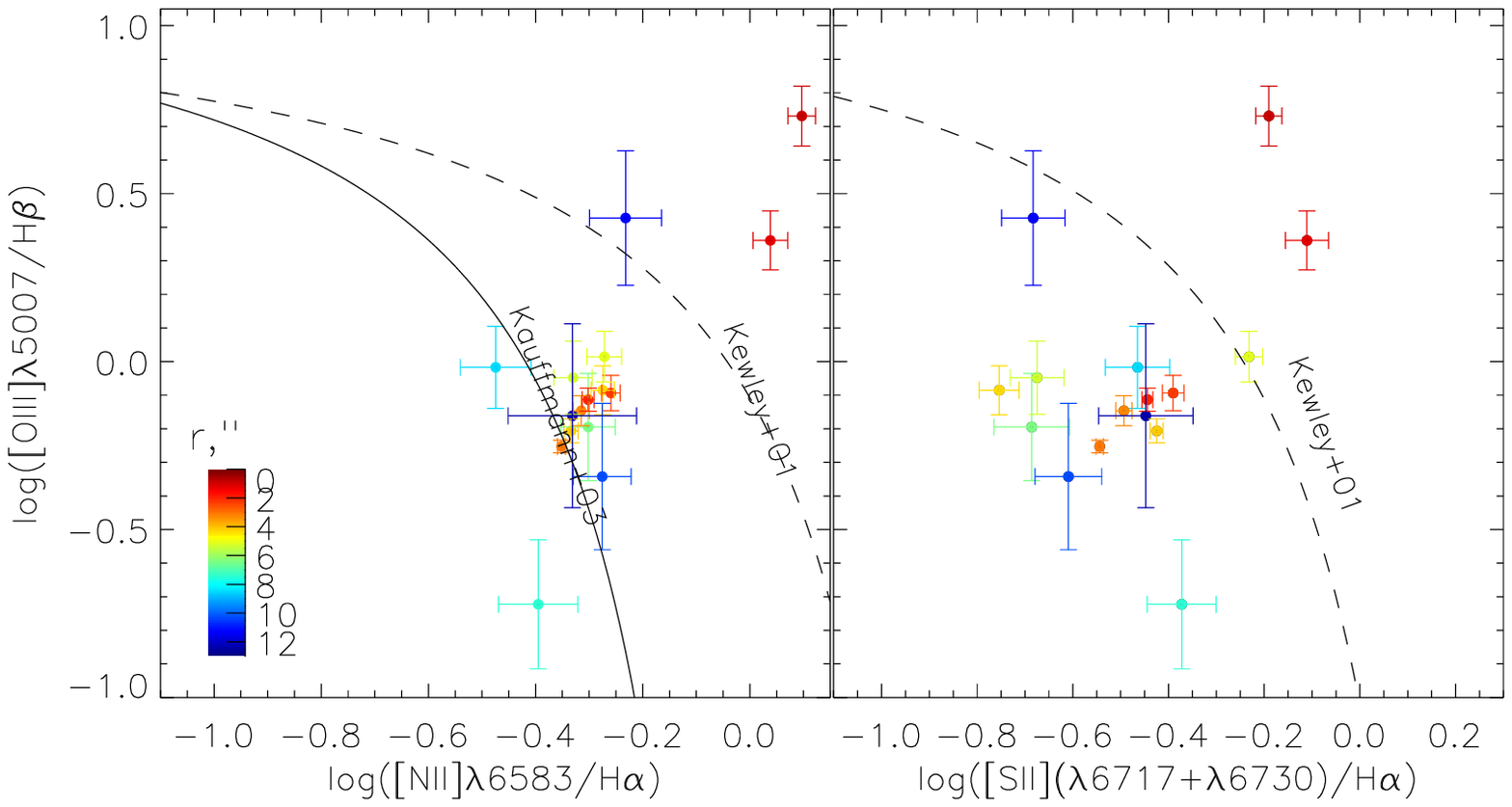}
\caption{J1237+39 diagnostic diagram of \OIIIHb\ versus  \NIIHa\  (left-hand panel) and \SIIHa\ (right-hand panel). Separating lines are the same as in Fig.~\ref{fig:bpt1117}.  Different symbol colours correspond to different radial distances according to the scale in the left-hand panel. 
}
\label{fig:bpt1237}
\end{figure*}

The BPT \citep*[after][]{BPT} diagnostic diagrams of the emission line ratios, derived from our long-slit data are presented in Figs.~\ref{fig:bpt1117} and \ref{fig:bpt1237}. The ???maximum starburst line??? \citep{Kewley2001} demarcation separates the regions, whose emission can be explained with photoionization by young massive stars  as a consequence of the ongoing star formation (\HII\ type) from those with the major contribution from other sources of excitation (shocks and Sy/LINER). The points lying between this  curve and the \citet{Kauffmann2003} curve  correspond to a composite excitation mechanism.

For both galaxies  we confirm the nuclear LINER-type  activity that was already mentioned in \citep{Wong2015} based on the SDSS nuclear spectra. Fig. \ref{fig:bpt1117} shows the changes   of gas excitation  with radial distances: the LINER  nucleus (orange dots), the composite mechanism in the star-forming ring in the \NIIHa\ versus \OIIIHb\ diagram $3''<r<7''$ (light-green dots in the diagrams), pure \HII\ type in the external disc ($7''<r<10''$, dark-green dots). This variations of the  forbidden-to-Balmer line ratios   are known as `AGN-HII mixing sequence' \citep{Kewley2006}. In J1117+51, this diagonal tracks across BPT diagrams could be related to domination of the star-forming processes and decrease of the AGN fraction of gas ionization, when the distance from the nucleus increases  \citep[see the model estimation by][]{Dickey2019}.  The influence of the AGN on gas ionization is observed only in the very center of the galaxy, where the blueshifted ionized gas outflow is detected (Sect.~\ref{sec:kin}).

In J1237+39, the   nuclear points also locate in the LINER region in the BPT diagrams (Fig.~\ref{fig:bpt1237}); however, without  a smooth radial changes of gas excitation properties: other circumnuclear points locate in the composite or \HII\ areas in the diagrams. Therefore, the weak  AGN in this galaxy has a very  small influence on the gas in the surrounding counter-rotating disc, even within the central kiloparsec region. This fact is in agreement with our previous conclusion about the absence of a nuclear outflow in this galaxy.

\section{Discussion}

In this paper,  we aimed to compare  the ionized gas properties and stellar kinematics  of four  `green valley' galaxies  with the cold gas mapped and  analysed by  \citet{Wong2015}. As is posited in the mentioned paper in these galaxies, \textit{`a kinetic process (possibly feedback from black hole activity) is driving the quick truncation of star formation in these systems, rather than a simple exhaustion of gas supply'}.

Spatial distribution of \HII\ correlates very well with the \HI\ data. First of all, for J0836+30 and J0900+46, the galaxies with the expelled \HI\ reservoirs, we confirm the absence of optical recombination lines in their discs. This fact indicates that either all the gas  was expelled (up to the detection limit) or fully ionized and warmed to $T_e>10^5$ K.  Also, in J1117+51 and J1237+39 we observe rotating  discs  on the same spatial scales with the \HI\ rotating structures. But unlike the \HI\ reservoir in J1117+51 that is blueshifted and lies beyond the optical radius of the galaxy, the ionized gas disc appears to lie within the optical disc and regularly rotates with a systemic velocity of $8240\km$.

Using a significantly  better angular resolution of optical data compared to those of 21 cm, we can study the gas kinematics in detail. Intriguing are the results for J1237+39: global counter rotation of gas and stars together with the velocity deviations  in the residual velocity field imply    that the history of this galaxy formation is much more complicated than that at the very start of the `quenching sequence'. Simulations and some observation studies  pointed out that major mergers are the most likely mechanism for generating a compact kinematically decoupled cores (KDC) in resulting galaxies, whereas  minor mergers or external accretion can produce dynamically cold large-scale gaseous  counter-rotating  discs and other structures like  bars or rings \citep{Tsatsi2015,Tapia2017,Pizzella2018}.

Moreover, the major merging usually creates dispersion-dominated slow rotators \citep[][and references therein]{Naab2014} which are different  from the rotation-dominated lenticulars  in our sample. Recently \citet*{Taylor2018} used cosmological simulations to demonstrate the formation of a large-scale counter-rotating gaseous disc via prolonged  cold-gas accretion from a cosmic web filament. Based on these simulations, in their next paper  \citet*{Taylor2019} have found that gas metallicity maps in the counter-rotating disc should have distinct areas of high and low metallicity separated by a sharp boundary. Unfortunately, the available observed data for the  J1237+39 galaxy are not enough to make a confident choice between minor merging and filamentary accretion  scenarios using the gas metallicity distribution. However,  the asymmetric features in the colour-index distribution accompanied by the non-circular gas motions in the inner 2--3 kpc  (see Sec.~\ref{sec:kin})  give additional arguments in favour of minor-merging hypothesis. It is possible that we directly observe a remnant of   the disturbed dwarf companion. New integral-field spectroscopic observations are essential to better understand the formation history of this galaxy.

The AGN feedback was proposed as the main mechanism of gas evacuation in the considered galaxies. Indeed, we see both the ionization and mechanical impacts of on the  AGN gas in J1117+51 and J1237+39, but the observed effects are  relatively modest and observed only in the cirumnuclear regions: non-thermal gas excitation and AGN-driven spatially unresolved gas outflow in  J1117+51. Therefore, the impact of the LINER nuclei on the extended gaseous  discs  in the  galaxies is negligible.

Traditionally,  ETGs include both elliptical and lenticular morphological types that might  cause misunderstanding. Indeed, 
\citet{Wong2015} have considered their sample galaxies as a major merging product, i.e., ellipticals. However,  J1117+51 and J1237+39 for which we obtained major axis stellar kinematics are rotation-dominated galaxies. Moreover, their SDSS images reveal numerous characteristics of morphological features of flat dymanically cold discs: bar or oval lens, rings, fragments of quasi-spiral structures. Therefore, at least a half of the sample are disc-sominated S0 or S0/a galaxies. Also, as we argue above, the gaseous disc in J1237+39 is of possible accretion origin.

All these lead to a suggestion that the \HI\ distribution and kinematics alone  cannot be an argument in favour of the stage of quenching, because  accretion is also possible. For better understanding the processes occurring in the `green valley' galaxies,  spatially-resolved optical spectroscopy data are necessary. Among local ETGs, a  comparison of \HI\ radio mapping with 3D-spectroscopy taken with the integral-field unit (IFU) was performed in the frameworks of the ATLAS$^{3D}$ survey \citep{Serra2012}. It seems interesting that the \HI\ counter-rotating disc galaxy J1237+39 lies in the same sequence as the ATLAS$^{3D}$ ETGs and ALFALFA gas-rich galaxies in the `\HI\ fraction -- colour' diagram \citep[see Fig.~9 in][]{Wong2015}. 
The HI-MaNGA project demonstrates a new step in this direction based on the huge sample of the MaNGA IFU survey \citep{Masters2019}. The detailed \HI\ radio synthesis observations of the selected MaNGA ETGs will be  a very interesting extension of the study presented in our current paper.


\section{Conclusion}

We have performed optical spectroscopic observations of four `green valley' ETGs that were already mapped in \HI\ and considered as  different stages of the star formation quenching under action of  the negative active nucleus feedback. We aimed to compare the \HI\ and \HII\ gas distributions and study the ionized gas kinematics with higher spatial resolution.  Main findings are the following:

\begin{enumerate}
    \item In two galaxies considered to be at the most advanced stages of quenching,  where \HI\ reservoirs are expelled, the similar picture is also observed in \HII: the ionized gas was not detected in the galaxy discs  outside the circumnuclear area. 
    \item In both galaxies having \HI\ counterparts of stellar discs, we also observed the ionized  gas discs with the dominating circular rotation. It is interesting that the \HII\ disc rotats regularly in J1117+51 that demonstrates the `HI displacement'. We suppose  that this difference in the \HI\ and \HII\ behaviour is caused by a relatively low resolution of the \HI\ data. 
    \item In both \HI-disc galaxies, the AGN affects the gas ionization and kinematics (i.e., the nuclear outflow is $\sim80\km$ in J1117+51).  However, the influence of the active nucleus was detected only in the central regions, unresolved in our observations ($<1$ kpc), we did not see an AGN feedback on larger scales.
    \item J1237+39  was considered to be at the first stage of quenching, however, its global gaseous disc (\HII+\HI) is in the counter rotation with the stellar one. Therefore, this galaxy may be on the way to start a new burst of star formation after minor merging or a gas accretion event.  The asymmetric structures in the SDSS colour-index distribution together with the non-circular gas motions give arguments in favour of recent minor merging in this galaxy.
\end{enumerate}

 \citet{Wong2015} suggested that \HI\ mapping are \textit{`excellent probes and measures of the quenching evolutionary stages'}. Our study of the same galaxy sample shows that a comparison  of 21-cm maps  with the spatially-resolved data on the properties of the ionized gas and stellar population is crucial to  understand the evolution paths of `green valley' galaxies. This comparison will be an interesting subject of current and new-generation IFU surveys. 
\section{Acknowledgements}
This study was supported by the Russian Science Foundation, project no. 17-12-01335 `Ionized gas in galaxy discs and beyond the optical radius' and based on observations conducted with the 6-m telescope of the Special Astrophysical Observatory of the Russian Academy of Sciences carried out with the financial support of the Ministry of Science and Higher Education of the Russian Federation. The authors thank
the anonymous referee for the detailed comments that helped
us to improve this manuscript and Gyula Jozsa  for his constructive comments. The research has made use of the NASA/IPAC Extragalactic Database (NED) which is operated by the Jet Propulsion Laboratory, California Institute of Technology, under contract with the National Aeronautics and Space Administration.

\label{lastpage}

\begin{thebibliography}{}
\makeatletter
\relax
\def\mn@urlcharsother{\let\do\@makeother \do\$\do\&\do\#\do\^\do\_\do\%\do\~}
\def\mn@doi{\begingroup\mn@urlcharsother \@ifnextchar [ {\mn@doi@}
  {\mn@doi@[]}}
\def\mn@doi@[#1]#2{\def\@tempa{#1}\ifx\@tempa\@empty \href
  {http://dx.doi.org/#2} {doi:#2}\else \href {http://dx.doi.org/#2} {#1}\fi
  \endgroup}
\def\mn@eprint#1#2{\mn@eprint@#1:#2::\@nil}
\def\mn@eprint@arXiv#1{\href {http://arxiv.org/abs/#1} {{\tt arXiv:#1}}}
\def\mn@eprint@dblp#1{\href {http://dblp.uni-trier.de/rec/bibtex/#1.xml}
  {dblp:#1}}
\def\mn@eprint@#1:#2:#3:#4\@nil{\def\@tempa {#1}\def\@tempb {#2}\def\@tempc
  {#3}\ifx \@tempc \@empty \let \@tempc \@tempb \let \@tempb \@tempa \fi \ifx
  \@tempb \@empty \def\@tempb {arXiv}\fi \@ifundefined
  {mn@eprint@\@tempb}{\@tempb:\@tempc}{\expandafter \expandafter \csname
  mn@eprint@\@tempb\endcsname \expandafter{\@tempc}}}

\bibitem[\protect\citeauthoryear{{Afanasiev} \& {Moiseev}}{{Afanasiev} \&
  {Moiseev}}{2011}]{SCORPIO2}
{Afanasiev} V.~L.,  {Moiseev} A.~V.,  2011, Baltic Astronomy, \href
  {http://adsabs.harvard.edu/abs/2011BaltA..20..363A} {20, 363}

\bibitem[\protect\citeauthoryear{{Baldry}, {Glazebrook}, {Brinkmann},
  {Ivezi{\'c}}, {Lupton}, {Nichol}  \& {Szalay}}{{Baldry}
  et~al.}{2004}]{Baldry04}
{Baldry} I.~K.,  {Glazebrook} K.,  {Brinkmann} J.,  {Ivezi{\'c}} {\v Z}.,
  {Lupton} R.~H.,  {Nichol} R.~C.,   {Szalay} A.~S.,  2004, \mn@doi [\apj]
  {10.1086/380092}, \href {http://adsabs.harvard.edu/abs/2004ApJ...600..681B}
  {600, 681}

\bibitem[\protect\citeauthoryear{{Baldry}, {Balogh}, {Bower}, {Glazebrook},
  {Nichol}, {Bamford}  \& {Budavari}}{{Baldry} et~al.}{2006}]{Baldry06}
{Baldry} I.~K.,  {Balogh} M.~L.,  {Bower} R.~G.,  {Glazebrook} K.,  {Nichol}
  R.~C.,  {Bamford} S.~P.,   {Budavari} T.,  2006, \mn@doi [\mnras]
  {10.1111/j.1365-2966.2006.11081.x}, \href
  {http://adsabs.harvard.edu/abs/2006MNRAS.373..469B} {373, 469}

\bibitem[\protect\citeauthoryear{{Baldwin}, {Phillips}  \&
  {Terlevich}}{{Baldwin} et~al.}{1981}]{BPT}
{Baldwin} J.~A.,  {Phillips} M.~M.,   {Terlevich} R.,  1981, \mn@doi [\pasp]
  {10.1086/130766}, \href {http://adsabs.harvard.edu/abs/1981PASP...93....5B}
  {93, 5}

\bibitem[\protect\citeauthoryear{{Begeman}}{{Begeman}}{1989}]{Begeman1989}
{Begeman} K.~G.,  1989, \aap, \href
  {http://adsabs.harvard.edu/abs/1989A%26A...223...47B} {223, 47}

\bibitem[\protect\citeauthoryear{{Blanton} \& {Moustakas}}{{Blanton} \&
  {Moustakas}}{2009}]{Blanton09}
{Blanton} M.~R.,  {Moustakas} J.,  2009, \mn@doi [\araa]
  {10.1146/annurev-astro-082708-101734}, \href
  {http://adsabs.harvard.edu/abs/2009ARA%26A..47..159B} {47, 159}

\bibitem[\protect\citeauthoryear{{Dickey}, {Geha}, {Wetzel}  \&
  {El-Badry}}{{Dickey} et~al.}{2019}]{Dickey2019}
{Dickey} C.,  {Geha} M.,  {Wetzel} A.,   {El-Badry} K.,  2019, arXiv e-prints,
  \href {http://ads.inasan.ru/abs/2019arXiv190201401D} {}

\bibitem[\protect\citeauthoryear{{Egorov}, {Lozinskaya}, {Moiseev}  \&
  {Smirnov-Pinchukov}}{{Egorov} et~al.}{2018}]{Egorov2018}
{Egorov} O.~V.,  {Lozinskaya} T.~A.,  {Moiseev} A.~V.,   {Smirnov-Pinchukov}
  G.~V.,  2018, \mn@doi [\mnras] {10.1093/mnras/sty1158}, \href
  {http://adsabs.harvard.edu/abs/2018MNRAS.478.3386E} {478, 3386}

\bibitem[\protect\citeauthoryear{{Erwin} \& {Sparke}}{{Erwin} \&
  {Sparke}}{1999}]{Erwin1999}
{Erwin} P.,  {Sparke} L.~S.,  1999, \mn@doi [\apj] {10.1086/312169}, \href
  {https://ui.adsabs.harvard.edu/abs/1999ApJ...521L..37E} {521, L37}

\bibitem[\protect\citeauthoryear{{Fasano}, {Poggianti}, {Couch}, {Bettoni},
  {Kj{\ae}rgaard}  \& {Moles}}{{Fasano} et~al.}{2000}]{Fasano2000}
{Fasano} G.,  {Poggianti} B.~M.,  {Couch} W.~J.,  {Bettoni} D.,
  {Kj{\ae}rgaard} P.,   {Moles} M.,  2000, \mn@doi [\apj] {10.1086/317047},
  \href {https://ui.adsabs.harvard.edu/abs/2000ApJ...542..673F} {542, 673}

\bibitem[\protect\citeauthoryear{{Finkelman}, {Moiseev}, {Brosch}  \&
  {Katkov}}{{Finkelman} et~al.}{2011}]{Finkelman2011}
{Finkelman} I.,  {Moiseev} A.,  {Brosch} N.,   {Katkov} I.,  2011, \mn@doi
  [\mnras] {10.1111/j.1365-2966.2011.19601.x}, \href
  {http://adsabs.harvard.edu/abs/2011MNRAS.418.1834F} {418, 1834}

\bibitem[\protect\citeauthoryear{{Fraternali} \& {Binney}}{{Fraternali} \&
  {Binney}}{2008}]{Fraternali2008}
{Fraternali} F.,  {Binney} J.~J.,  2008, \mn@doi [\mnras]
  {10.1111/j.1365-2966.2008.13071.x}, \href
  {https://ui.adsabs.harvard.edu/abs/2008MNRAS.386..935F} {386, 935}

\bibitem[\protect\citeauthoryear{{Hubble}}{{Hubble}}{1926}]{Hubble1926}
{Hubble} E.~P.,  1926, \mn@doi [\apj] {10.1086/143018}, \href
  {https://ui.adsabs.harvard.edu/abs/1926ApJ....64..321H} {64, 321}

\bibitem[\protect\citeauthoryear{{Jin} et~al.,}{{Jin} et~al.}{2016}]{Jin2016}
{Jin} Y.,  et~al., 2016, \mn@doi [\mnras] {10.1093/mnras/stw2055}, \href
  {http://ads.inasan.ru/abs/2016MNRAS.463..913J} {463, 913}

\bibitem[\protect\citeauthoryear{{Katkov}, {Kniazev}  \& {Sil'chenko}}{{Katkov}
  et~al.}{2015}]{Katkov2015}
{Katkov} I.~Y.,  {Kniazev} A.~Y.,   {Sil'chenko} O.~K.,  2015, \mn@doi [\aj]
  {10.1088/0004-6256/150/1/24}, \href
  {https://ui.adsabs.harvard.edu/abs/2015AJ....150...24K} {150, 24}

\bibitem[\protect\citeauthoryear{{Kauffmann} et~al.,}{{Kauffmann}
  et~al.}{2003}]{Kauffmann2003}
{Kauffmann} G.,  et~al., 2003, \mn@doi [\mnras]
  {10.1111/j.1365-2966.2003.07154.x}, \href
  {http://adsabs.harvard.edu/abs/2003MNRAS.346.1055K} {346, 1055}

\bibitem[\protect\citeauthoryear{{Kewley}, {Dopita}, {Sutherland}, {Heisler}
  \& {Trevena}}{{Kewley} et~al.}{2001}]{Kewley2001}
{Kewley} L.~J.,  {Dopita} M.~A.,  {Sutherland} R.~S.,  {Heisler} C.~A.,
  {Trevena} J.,  2001, \mn@doi [\apj] {10.1086/321545}, \href
  {http://adsabs.harvard.edu/abs/2001ApJ...556..121K} {556, 121}

\bibitem[\protect\citeauthoryear{{Kewley}, {Groves}, {Kauffmann}  \&
  {Heckman}}{{Kewley} et~al.}{2006}]{Kewley2006}
{Kewley} L.~J.,  {Groves} B.,  {Kauffmann} G.,   {Heckman} T.,  2006, \mn@doi
  [\mnras] {10.1111/j.1365-2966.2006.10859.x}, \href
  {http://adsabs.harvard.edu/abs/2006MNRAS.372..961K} {372, 961}

\bibitem[\protect\citeauthoryear{{Koleva}, {Prugniel}, {Bouchard}  \&
  {Wu}}{{Koleva} et~al.}{2009}]{ULySS}
{Koleva} M.,  {Prugniel} P.,  {Bouchard} A.,   {Wu} Y.,  2009, \mn@doi [\aap]
  {10.1051/0004-6361/200811467}, \href
  {http://adsabs.harvard.edu/abs/2009A%26A...501.1269K} {501, 1269}

\bibitem[\protect\citeauthoryear{{Martin} et~al.,}{{Martin}
  et~al.}{2007}]{2007ApJS..173..342M}
{Martin} D.~C.,  et~al., 2007, \mn@doi [\apjs] {10.1086/516639}, \href
  {http://adsabs.harvard.edu/abs/2007ApJS..173..342M} {173, 342}

\bibitem[\protect\citeauthoryear{{Masters} et~al.,}{{Masters}
  et~al.}{2019}]{Masters2019}
{Masters} K.~L.,  et~al., 2019, arXiv e-prints, \href
  {https://ui.adsabs.harvard.edu/abs/2019arXiv190105579M} {p. arXiv:1901.05579}

\bibitem[\protect\citeauthoryear{{Moiseev}}{{Moiseev}}{2014}]{Moiseev2014}
{Moiseev} A.~V.,  2014, \mn@doi [Astrophysical Bulletin]
  {10.1134/S1990341314010015}, \href
  {http://adsabs.harvard.edu/abs/2014AstBu..69....1M} {69, 1}

\bibitem[\protect\citeauthoryear{{Moiseev}}{{Moiseev}}{2015}]{Moiseev2015}
{Moiseev} A.~V.,  2015, \mn@doi [Astrophysical Bulletin]
  {10.1134/S1990341315040112}, \href
  {http://adsabs.harvard.edu/abs/2015AstBu..70..494M} {70, 494}

\bibitem[\protect\citeauthoryear{{Moiseev} \& {Egorov}}{{Moiseev} \&
  {Egorov}}{2008}]{MoiseevEgorov2008}
{Moiseev} A.~V.,  {Egorov} O.~V.,  2008, \mn@doi [Astrophysical Bulletin]
  {10.1134/S1990341308020089}, \href
  {http://adsabs.harvard.edu/abs/2008AstBu..63..181M} {63, 181}

\bibitem[\protect\citeauthoryear{{Moiseev}, {Vald{\'e}s}  \&
  {Chavushyan}}{{Moiseev} et~al.}{2004}]{Moiseev2004}
{Moiseev} A.~V.,  {Vald{\'e}s} J.~R.,   {Chavushyan} V.~H.,  2004, \mn@doi
  [\aap] {10.1051/0004-6361:20040045}, \href
  {http://adsabs.harvard.edu/abs/2004A%26A...421..433M} {421, 433}

\bibitem[\protect\citeauthoryear{{Naab} et~al.,}{{Naab}
  et~al.}{2014}]{Naab2014}
{Naab} T.,  et~al., 2014, \mn@doi [\mnras] {10.1093/mnras/stt1919}, \href
  {https://ui.adsabs.harvard.edu/abs/2014MNRAS.444.3357N} {444, 3357}

\bibitem[\protect\citeauthoryear{{Pizzella}, {Morelli}, {Coccato}, {Corsini},
  {Dalla Bont{\`a}}, {Fabricius}  \& {Saglia}}{{Pizzella}
  et~al.}{2018}]{Pizzella2018}
{Pizzella} A.,  {Morelli} L.,  {Coccato} L.,  {Corsini} E.~M.,  {Dalla
  Bont{\`a}} E.,  {Fabricius} M.,   {Saglia} R.~P.,  2018, \mn@doi [\aap]
  {10.1051/0004-6361/201731712}, \href
  {http://adsabs.harvard.edu/abs/2018A%26A...616A..22P} {616, A22}

\bibitem[\protect\citeauthoryear{{Querejeta} et~al.,}{{Querejeta}
  et~al.}{2015}]{Querejeta2015}
{Querejeta} M.,  et~al., 2015, \mn@doi [\aap] {10.1051/0004-6361/201526354},
  \href {https://ui.adsabs.harvard.edu/abs/2015A&A...579L...2Q} {579, L2}

\bibitem[\protect\citeauthoryear{{Serra} et~al.,}{{Serra}
  et~al.}{2012}]{Serra2012}
{Serra} P.,  et~al., 2012, \mn@doi [\mnras] {10.1111/j.1365-2966.2012.20219.x},
  \href {https://ui.adsabs.harvard.edu/abs/2012MNRAS.422.1835S} {422, 1835}

\bibitem[\protect\citeauthoryear{{Sil'chenko}, {Moiseev}  \&
  {Afanasiev}}{{Sil'chenko} et~al.}{2009}]{Silchenko2009}
{Sil'chenko} O.~K.,  {Moiseev} A.~V.,   {Afanasiev} V.~L.,  2009, \mn@doi
  [\apj] {10.1088/0004-637X/694/2/1550}, \href
  {http://ads.inasan.ru/abs/2009ApJ...694.1550S} {694, 1550}

\bibitem[\protect\citeauthoryear{{Sil'chenko}, {Proshina}, {Shulga}  \&
  {Koposov}}{{Sil'chenko} et~al.}{2012}]{Silchenko2012}
{Sil'chenko} O.~K.,  {Proshina} I.~S.,  {Shulga} A.~P.,   {Koposov} S.~E.,
  2012, \mn@doi [\mnras] {10.1111/j.1365-2966.2012.21990.x}, \href
  {https://ui.adsabs.harvard.edu/abs/2012MNRAS.427..790S} {427, 790}

\bibitem[\protect\citeauthoryear{{Sil'chenko}, {Moiseev}  \&
  {Egorov}}{{Sil'chenko} et~al.}{2019}]{Silchenko2019}
{Sil'chenko} O.~K.,  {Moiseev} A.~V.,   {Egorov} O.~V.,  2019, \apjs, in press,
  \href {https://ui.adsabs.harvard.edu/abs/2019arXiv190707261S} {p.
  arXiv:1907.07261}

\bibitem[\protect\citeauthoryear{{Strateva} et~al.,}{{Strateva}
  et~al.}{2001}]{Strateva}
{Strateva} I.,  et~al., 2001, \mn@doi [\aj] {10.1086/323301}, \href
  {http://adsabs.harvard.edu/abs/2001AJ....122.1861S} {122, 1861}

\bibitem[\protect\citeauthoryear{{Tapia}, {Eliche-Moral}, {Aceves},
  {Rodr{\'\i}guez-P{\'e}rez}, {Borlaff}  \& {Querejeta}}{{Tapia}
  et~al.}{2017}]{Tapia2017}
{Tapia} T.,  {Eliche-Moral} M.~C.,  {Aceves} H.,  {Rodr{\'\i}guez-P{\'e}rez}
  C.,  {Borlaff} A.,   {Querejeta} M.,  2017, \mn@doi [\aap]
  {10.1051/0004-6361/201628821}, \href
  {https://ui.adsabs.harvard.edu/abs/2017A&A...604A.105T} {604, A105}

\bibitem[\protect\citeauthoryear{{Taylor}, {Federrath}  \&
  {Kobayashi}}{{Taylor} et~al.}{2018}]{Taylor2018}
{Taylor} P.,  {Federrath} C.,   {Kobayashi} C.,  2018, \mn@doi [\mnras]
  {10.1093/mnras/sty1439}, \href
  {https://ui.adsabs.harvard.edu/abs/2018MNRAS.479..141T} {479, 141}

\bibitem[\protect\citeauthoryear{{Taylor}, {Kobayashi}  \&
  {Federrath}}{{Taylor} et~al.}{2019}]{Taylor2019}
{Taylor} P.,  {Kobayashi} C.,   {Federrath} C.,  2019, \mn@doi [\mnras]
  {10.1093/mnras/stz630}, \href
  {https://ui.adsabs.harvard.edu/abs/2019MNRAS.485.3215T} {485, 3215}

\bibitem[\protect\citeauthoryear{{Tsatsi}, {Macci{\`o}}, {van de Ven}  \&
  {Moster}}{{Tsatsi} et~al.}{2015}]{Tsatsi2015}
{Tsatsi} A.,  {Macci{\`o}} A.~V.,  {van de Ven} G.,   {Moster} B.~P.,  2015,
  \mn@doi [\apj] {10.1088/2041-8205/802/1/L3}, \href
  {https://ui.adsabs.harvard.edu/abs/2015ApJ...802L...3T} {802, L3}

\bibitem[\protect\citeauthoryear{{Wilman}, {Oemler}, {Mulchaey}, {McGee},
  {Balogh}  \& {Bower}}{{Wilman} et~al.}{2009}]{Wilman2009}
{Wilman} D.~J.,  {Oemler} A. J.,  {Mulchaey} J.~S.,  {McGee} S.~L.,  {Balogh}
  M.~L.,   {Bower} R.~G.,  2009, \mn@doi [\apj] {10.1088/0004-637X/692/1/298},
  \href {https://ui.adsabs.harvard.edu/abs/2009ApJ...692..298W} {692, 298}

\bibitem[\protect\citeauthoryear{{Wong}, {Schawinski}, {J{\'o}zsa}, {Urry},
  {Lintott}, {Simmons}, {Kaviraj}  \& {Masters}}{{Wong}
  et~al.}{2015}]{Wong2015}
{Wong} O.~I.,  {Schawinski} K.,  {J{\'o}zsa} G.~I.~G.,  {Urry} C.~M.,
  {Lintott} C.~J.,  {Simmons} B.~D.,  {Kaviraj} S.,   {Masters} K.~L.,  2015,
  \mn@doi [\mnras] {10.1093/mnras/stu2724}, \href
  {http://adsabs.harvard.edu/abs/2015MNRAS.447.3311W} {447, 3311}

\makeatother
\end{thebibliography}
\end{document}